\documentclass[11pt]{article}

\usepackage{amsmath}
\usepackage{amssymb}
\usepackage{latexsym}
\usepackage{graphicx}
\usepackage{epsfig}
\usepackage{stmaryrd}
\usepackage{framed}

\newtheorem{thm}{\bf Theorem}[section]
\newtheorem{lem}[thm]{\bf Lemma}        
\newtheorem{prop}[thm]{\bf Proposition}  
\newtheorem{cor}[thm]{\bf Corollary}

\newtheorem{expl}[thm]{\bf Example}       
\newtheorem{remark}[thm]{\bf Remark}

\newcommand{\op}{\mathrm{op}}

\def\Bbox{
{\unskip\nobreak\hfil\penalty50
\hskip1em\hbox{}\nobreak\hfil{\lower .5pt \hbox{$\Box$}}
\parfillskip=0pt \finalhyphendemerits=0 \par}
}

\def\eop{
\ifmmode {\hbox{\Bbox}} \else \Bbox \fi
}

\def\bbox{
\ifmmode {\hbox{\bbox}} \else \Bbox \fi
}

\begin{document}

\title{{\bf A representation theorem for stratified complete lattices}}


\author{Zolt\'an \'Esik\thanks{Partially supported by 
grant no. ANN 110883 from the National Foundation of Hungary for 
Scientific Research.}\\
Dept. of Computer Science\\
University of Szeged\\
Hungary}

\maketitle

\begin{abstract}
We consider complete lattices equipped with preorderings indexed by 
the ordinals less than a given (limit) ordinal subject to certain
axioms. These structures, called stratified complete lattices, 
and weakly monotone functions over them, provide a framework 
for solving fixed point equations involving non-monotone operations 
such as negation or complement, and have been used to give semantics 
to logic programs with negation.  

More precisely, we consider stratified complete lattices subject to 
two slightly different systems 
of axioms defining `models' and `strong models'. We prove that a stratified 
complete lattice is a model iff it is isomorphic to the stratified 
complete lattice determined 
by the limit of an inverse system of complete lattices with `locally completely 
additive' projections. Moreover, we prove that a stratified 
complete lattice is a strong model iff it is isomorphic to the 
stratified complete lattice determined 
by the limit of an inverse system of complete lattices with completely 
additive projections.

We use the inverse limit representation   
to give alternative proofs of some recent results and to derive some new 
ones for models and strong models. 
In particular, we use the representation theorem to prove that 
every model gives rise to another complete lattice structure,
which in limit models corresponds to the lexicographic order. 
Moreover, we prove that the set of all fixed points of a weakly 
monotone function over a model, equipped with the
new ordering, is a complete lattice. We also consider symmetric models that 
satisfy, together with each axiom, the dual axiom, and use the inverse limit
representation to prove that every strong model is symmetric. 
\end{abstract}

\section{Introduction}

The motivation for this paper comes from logic programming. 
The most commonly accepted semantic model of a  logic 
program with negation is the well-founded model, c.f.
\cite{Gelder,Prz}. The well-founded approach to the semantics of
logic programs with negation is based on a three-valued
(or sometimes four-valued) logic and describes the meaning of a 
logic program as the least fixed point of the so-called 
`stable operator' canonically associated with 
the program with respect to the information, or knowledge, 
or Fitting ordering \cite{Fitting} of interpretations.  
The well-founded approach to logic programming has led to the development of a deep abstract 
fixed point theory for non-monotone functions with several applications 
 beyond logic programming, 
see \cite{Deneckeretal1,Deneckeretal,Fitting,Vennekensetal}
for a sampling of articles covering such results. 

Another approach to the semantics of logic programs with negation,
using an infinite supply of truth values, 
was introduced in \cite{RW}. The development of a fixed point theory 
underlying this approach has recently been undertaken in \cite{Esax,ERfp,ERwollic}. 
This fixed point theory has been applied to higher-order 
logic programs with negation \cite{CERhigher}
and to Boolean context-free languages \cite{ERwollic}.
(Boolean context-free languages were introduced in \cite{Okhotin}
and are closely related to some of the language equations 
in \cite{Leiss}.) 

The structures studied in this novel fixed point theory are 
stratified complete lattices, i.e., complete lattices $(L,\leq)$, 
equipped with a family of preorderings $\sqsubseteq_\alpha$, indexed 
by the ordinals $\alpha$ strictly less than a fixed nonzero ordinal $\kappa$, which 
without loss of generality can be taken to be a limit ordinal. 
In \cite{ERfp,ERwollic}, several systems of axioms have been introduced.
Some of the results, such as the `Lattice Theorem' or 
the `Fixed Point Theorem' of \cite{ERfp}, were proved for a weaker 
class of models, whereas 
some others, such as the `Model Intersection Theorem'
of \cite{ERwollic}, were established for stronger classes of models. 
The Lattice Theorem asserts that every model $L$ of the axioms can be 
equipped with another complete lattice ordering $\sqsubseteq$ 
by defining $x \sqsubseteq y$ iff either $x= y$, or there is some $\alpha < \kappa$
with $x \sqsubset_\alpha y$ (i.e., $x \sqsubseteq_\alpha y$
but $y \not\sqsubseteq_\alpha x$). The Fixed Point Theorem states that 
certain weakly monotone functions $L \to L$ have least fixed points
w.r.t. the ordering $\sqsubseteq$.  

In this paper, we deal with two systems of axioms 
introduced in \cite{ERfp,ERwollic} that seem to be the
most relevant to applications. In the stratified complete lattices 
satisfying these systems of axioms, called models and strong models, resp., 
the preorderings $\sqsubseteq_\alpha$, $\alpha < \kappa$, are completely determined by 
the complete lattice order $\leq$ and the equivalence relations 
$=_\alpha$  corresponding to the preorderings $\sqsubseteq_\alpha$.

The main results of the paper are:
\begin{itemize}
\item Every model $L$ is isomorphic to the stratified complete lattice
determined by an inverse limit of complete lattices with locally completely 
additive projections, cf. Theorem~\ref{thm-rep}.
\item Every strong model $L$ is isomorphic to the stratified 
complete lattice determined by 
an inverse limit of complete lattices with completely 
additive projections, cf. Corollary~\ref{cor-rep}.
\item A general result (Theorem~\ref{thm-general}) based on the above representation theorems
implying the Lattice Theorem, the Fixed Point Theorem, and 
the fact that for every model $L$ and weakly monotone 
function $f: L \to L$ w.r.t. $\sqsubseteq$,
the fixed points of $L$ form a complete lattice w.r.t. the ordering 
$\sqsubseteq$.
\end{itemize}

The paper is organized as follows. 
In Section~\ref{sec1}, we define models and strong models 
by means of two systems of axioms originating from \cite{ERfp}.
We discuss some examples including a
 model that was used in \cite{RW} to give semantics to logic programs 
with negation, and the product models from \cite{ERfp},  
constructed from a well-ordered collection of complete lattices.   
Then, in Section~\ref{sec-inverselimits}, we study inverse systems
$h^\alpha_\beta: L_\alpha \to L_\beta$, $\beta < \alpha < \kappa$, 
where each $L_\alpha$ is a complete lattice and the functions $h^\alpha_\beta$
are projections, sometimes also locally completely additive 
(as defined in the paper), or 
completely additive. Then, in Section~\ref{sec-inverse limit models}, we show that 
if the functions $h^\alpha_\beta$ in the inverse system are locally completely additive
projections, then the limit of the inverse system gives rise to a model, which is a
strong model if the functions $h^\alpha_\beta$ are completely additive projections. 
In Section~\ref{sec-cons} we discuss several consequences of the axioms 
and provide a brief analysis of the interconnection between them. 
In Section~\ref{sec-alternative}, we provide alternative 
axiomatizations of both models and strong models using a family of functions $|_\alpha$
instead of the relations $\sqsubseteq_\alpha$, $\alpha < \kappa$. 
Then, in Section~\ref{sec-rep}, we use the properties established 
in Section~\ref{sec-cons} and Section~\ref{sec-alternative} to prove 
the Representation Theorem (Theorem~\ref{thm-rep}) 
and its corollaries showing that every model is 
isomorphic to the limit model
obtained from an inverse system of complete lattices 
with locally completely additive projections, 
and every strong model is isomorphic to the limit 
model determined by an inverse system of complete lattices 
with completely additive projections. 
In limit models, the ordering $\sqsubseteq$
corresponds to the lexicographic ordering. 
Actually we show that the stratified complete lattice determined by an inverse system 
of complete lattices $L_\alpha$, $\alpha < \kappa$, with projections 
$h^\alpha_\beta: L_\alpha \to L_\beta$, $\beta < \alpha < \kappa$,
is a model (strong model, resp.) iff each projection $h^\alpha_\beta$ 
is locally completely additive (completely additive, resp.). 
In Section~\ref{sec-further}, we use the Representation Theorem to establish 
Theorem~\ref{thm-general},  which in turn implies the Lattice Theorem
and the Fixed Point Theorem. In fact, 
Theorem~\ref{thm-general} is used to establish 
a novel result to the effect that the fixed points of a weakly 
monotone function over a model form a complete lattice
w.r.t. the ordering $\sqsubseteq$,
see Corollary~\ref{cor-latticefp}. Section~\ref{sec-symmetric} is devoted 
to symmetric models and strong symmetric models satisfying together with each 
axiom the dual axiom. We prove that a model is strong iff it is symmetric 
iff it is a strong symmetric model, whereas there is a model that is 
not symmetric. 
The paper ends with some concluding remarks.

\section{Models and examples}
\label{sec1} 

In this section, we introduce axioms for the structures we are going to 
discuss throughout the paper. We will also provide some examples and 
a construction. For unexplained notions regarding lattices we refer to 
\cite{Daveyetal}.

Suppose that $\kappa$ is a fixed limit ordinal. 
We will be considering structures of the sort 
$L = (L,\leq,(\sqsubseteq_\alpha)_{\alpha < \kappa})$,
called \emph{stratified complete lattices},  
such that $(L,\leq)$ is a complete lattice (with 
bottom and top elements $\bot$ and $\top$, resp.), 
and for each $\alpha < \kappa$, $\sqsubseteq_\alpha$ 
is a preordering of $L$. 

Our stratified complete lattices will satisfy the 
following axioms, where for each $\alpha$, 
$=_\alpha$ denotes the equivalence relation determined by
$\sqsubseteq_\alpha$.  
{\em 
\begin{itemize}
\item A1. For all $\alpha < \beta < \kappa$, $\sqsubseteq_\beta$ is included in $=_\alpha$,
so that if $x \sqsubseteq_\beta y$ then $x =_\alpha y$.
\item A2. The intersection of all the relations $=_\alpha$ for $\alpha < \kappa$ 
is the identity relation, so that if $x =_\alpha y$ for all $\alpha < \kappa$,
then $x = y$. 
\item A3. For all $x$ and $\alpha < \kappa$ there exists $y$ such that  $x=_\alpha y$
and for all $z$, if $x \sqsubseteq_\alpha  z$ then $y \leq z$.
\end{itemize}
}
It follows from the first two axioms that the intersection of all relations $\sqsubseteq_\alpha$,
$\alpha < \kappa$,  is also the identity relation.
It is clear that the element $y$ in A3 is uniquely determined by $x$ and $\alpha$ 
and also satisfies $y \sqsubseteq_\alpha z$ whenever $x \sqsubseteq_\alpha z$. 
We will denote it by $x|_\alpha$.  
{\em 
\begin{itemize}
\item
A4. For all $\alpha$ with $\alpha < \kappa$ and $x_i$ and $y$ with 
    $x_i =_\alpha y$, $i \in I$, where $I$ is any nonempty index set,
    it holds that $\bigvee_{i \in I}x_i =_\alpha y$.
\item
A5. For all $x,y$ and $\alpha < \kappa$, if $x \leq y$ then $x|_\alpha \leq y|_\alpha$.
\item
A6. For all $x,y$ and $\alpha < \kappa$, if $x \leq y$ and $x =_\beta y$ for all $\beta < \alpha$, then 
$x \sqsubseteq_\alpha y$.
\end{itemize}
}
A stratified complete lattice satisfying the above axioms A1--A6 will be called a 
 \emph{model}, for short. 

Sometimes we will require a stronger variant of A4. 
{\em 
\begin{itemize}
\item
A4$^*$. For all $\alpha$ with $\alpha < \kappa$ and $x_i,y_i$ with 
    $x_i =_\alpha y_i$, $i \in I$, where $I$ is any (nonempty) index set,
    it holds that $\bigvee_{i \in I}x_i =_\alpha \bigvee_{i \in I} y_i$.
\end{itemize} 
}
Models satisfying A4$^*$ will be called \emph{strong}. 
We will discuss several consequences of the axioms in Section~\ref{sec-cons}.

The following motivating example is from \cite{ERfp,RW}.
Consider the following linearly ordered set $V= V_\kappa$ of truth values:
$$F_0 < F_1< \ldots < F_\alpha < \ldots < 0 < \ldots < T_\alpha < \ldots < T_1 < T_0,$$
where $\alpha$ ranges over the ordinals strictly less than $\kappa$.
Let $Z$ denote a nonempty set of (propositional) variables and consider the set $L = V^Z$, 
equipped with the pointwise ordering. Thus, for all $f,g \in L$,
$f \leq g$ iff $f(z) \leq g(z)$ for all $z \in Z$. Then $(L,\leq)$ is a
complete lattice. 
For each $f,g \in L$ and $\alpha < \kappa$, define $f \sqsubseteq_\alpha g$ 
iff for all $z\in Z$,
\begin{itemize}
\item $\forall \beta < \alpha$ $(f(z) = F_\beta \Leftrightarrow  g(z) = F_\beta$\ $\wedge$\ 
   $f(z) = T_\beta \Leftrightarrow g(z) = T_\beta)$,
\item  $g(z) = F_\alpha \Rightarrow f(z) = F_\alpha \wedge f(x) = T_\alpha \Rightarrow 
g(z) = T_\alpha$. 
\end{itemize}
Then $L$ is a strong model. When $f \in L$ and $\alpha < \kappa$, then for all $z \in Z$,
$f|_\alpha(z) = f(z)$ if $f(z)$ is in the set $\{F_\beta,T_\beta : \beta \leq \alpha\}$,
and $f|_\alpha(z) = F_{\alpha + 1}$, otherwise. 
For $\kappa$ being the least uncountable ordinal $\Omega$, 
this example was used in \cite{RW} to give semantics to possibly countably 
infinite propositional logic programs involving negation. The idea is to associate with a logic 
program $P$ over $Z$ a function $f_P: V_\Omega^Z \to V_\Omega^Z$, and to define the semantics 
of $P$ as the unique least fixed point of $f_P$ with respect to a new ordering 
$\sqsubseteq$, canonically defined for interpretations $I,J$ in $V_\Omega^Z$ 
by $I \sqsubseteq J$ iff $I = J$ or there is some $\alpha < \Omega$ with 
$f_P(I) \sqsubset_\alpha f_P(J)$ (i.e., $f_P(I) \sqsubseteq_\alpha f_P(J)$ but 
$f_P(J) \not \sqsubseteq_\alpha f_P(I)$). The function $f_P$ is not necessarily 
monotone with respect to $\sqsubseteq$. 
It is argued in \cite{RW} that the semantics corresponds to the view of negation as failure. 
See Example~\ref{expl-logic program} for more details.
For an extension to higher order logic programs, see \cite{CERhigher}. 

In particular, $Z$ can be chosen to be a singleton set.
It follows that $V_\kappa$ is itself a strong model with the relations $\sqsubseteq_\alpha$,
$\alpha < \kappa$, 
defined by $x \sqsubseteq_\alpha y$ iff $x = y$ 
or $x,y \in \{F_\gamma, T_\gamma : \gamma \geq \alpha\} \cup \{0\}$
such that 
if $x = T_\alpha$ then $y = T_\alpha$ and if $y = F_\alpha$ then $x = F_\alpha$. 

We now describe a  construction of models.

\begin{expl}
{\rm 
\cite{ERfp} 
Suppose that $(L_\alpha,\leq)$ is a complete lattice with least
and greatest elements $\bot_\alpha$ and $\top_\alpha$, for all $\alpha < \kappa$.
Let $L$ be the direct product $\prod_{\alpha < \kappa} L_\alpha$, ordered pointwise,
so that for all $x = (x_\alpha)_{\alpha < \kappa}$ and $y = (y_\alpha)_{\alpha < \kappa}$ in $L$, 
$x \leq y$ iff $x_\alpha \leq y_\alpha$ for all $\alpha < \kappa$.
It is well-known that $L$ is also a complete lattice in which both the infimum 
and the supremum of any set is formed pointwise. For each $\alpha < \kappa$ 
and $x$ and $y$ as above, define $x \sqsubseteq_\alpha y$ iff $x_\alpha \leq y_\alpha$ 
and $x_\beta = y_\beta$ for all $\beta < \alpha$. Then $(L,\leq,(\sqsubseteq_\alpha)_{\alpha < \kappa})$ 
is a strong model, called a \emph{product model}.  In particular, for all $x = (x_\gamma)_{\gamma < \kappa}$ and $\alpha < \kappa$, 
$x|_\alpha = (y_\gamma)_{\gamma < \kappa}$ is given by 
$y_\gamma = x_\gamma$ for all $\gamma \leq \alpha$, and $y_\gamma = \bot_\gamma$ 
for all $\gamma$ with $\alpha < \gamma < \kappa$.
}
\end{expl} 

\begin{remark}
The axioms A1--A6 are from \cite{ERfp} and \cite{ERwollic}. Actually A3 is a weaker version 
of the corresponding axiom in \cite{ERfp} that we will denote A3$^*$. 
(Axiom A3$^*$ will be recalled and established in all models in Proposition~\ref{prop-p1}.) 

Several results for models have been reported in \cite{ERfp} and \cite{ERwollic}, 
albeit under varying assumptions. For example, the `Lattice Theorem' and the `Fixed Point Theorem' 
were proved in \cite{ERfp} using axioms A1, A2, A3$^*$ and A4, while the `Model Intersection Theorem' 
of \cite{ERwollic} was proved using axioms A1--A6, with A3 being replaced by the stronger A3$^*$.
However, all of the stratified complete lattices used in applications (logic programming, Boolean
context-free grammars) in \cite{CERhigher,ERfp} have thus far been models of the axioms A1--A6. 
\end{remark}

\section{Inverse limits}
\label{sec-inverselimits}

In this section, we recall the notion of inverse 
systems and limits of inverse systems of complete lattices. 
Inverse limits will be used to construct further models
of the axioms. We will make use of the following concept. 

Suppose that $L = (L,\leq)$ and $L' = (L',\leq)$ are complete lattices.
We say that $h: L' \to L$ \emph{preserves all infima}
if $h(\bigwedge Y) = \bigwedge h(Y)$ for all $Y \subseteq L$.  
Similarly, we say that $k: L \to L'$ \emph{preserves all suprema}, or that $k$ is 
\emph{completely additive}, if $k(\bigvee X) = \bigvee k(X)$ for all $X \subseteq L$.
It is clear that if $h: L' \to L$ preserves all infima,
then it is monotone and preserves the greatest element.
If $h$ is additionally surjective, then it preserves the least element. 
Similar facts hold for functions preserving all suprema.

Suppose that $L$ and $L'$ are complete lattices and $h: L' \to L$ and $k: L \to L'$ 
are monotone functions. We say that $(h,k)$ is a \emph{(monotone) Galois connection} 
\cite{Daveyetal}
(with $h$ being the upper and $k$ being the lower adjoint)
if the identity function on $L$ is less than or equal to $h \circ k : L \to L$ 
and $k \circ h : L' \to L'$ is less than or equal to the identity function on $L'$ 
with respect to the pointwise ordering of functions. It is known, 
cf. \cite{Daveyetal}, that for complete lattices $L$ and $L'$ and functions 
$h: L' \to L$ and $k: L \to L'$, $(h,k)$ is a Galois connection 
iff $h$ preserves all infima and $k$ preserves all suprema. 
Moreover, we say that $(h,k)$ is a \emph{projection-embedding pair} \cite{Scott} if 
$h \circ k : L \to L$ is the identity function on $L$ 
and $k \circ h : L' \to L'$ is less than or equal to the identity function on $L'$ 
with respect to the pointwise ordering of functions. 
Thus, a projection-embedding pair is a Galois connection. 

Suppose that $(h,k)$ is a Galois connection between complete lattices 
as above. If $(h,k)$ is a projection embedding pair, then $h$ is 
clearly surjective and $k$ is injective. Conversely, if 
$h$ is surjective or $k$ is injective, then $(h,k)$ 
is a projection-embedding pair (also called a Galois insertion). 
It is also clear that and $h$ uniquely determines $k$ and vice versa. 
Indeed, for each $x \in L$, $k(x)$ is the least element $y$ of $L'$ 
with $x \leq h(y)$. And for each $y \in L'$, $h(y)$ is the greatest 
$x \in L$ with $k(x) \leq y$.

We call $h: L' \to L$ a \emph{projection} if it is monotone and there is a 
corresponding embedding $L \to L'$ (which is then uniquely determined),
and call $k: L \to L'$ an \emph{embedding} if it is monotone 
and there is a corresponding projection
$L' \to L$. A well-known useful fact is that any composition of projections is a projection
and corresponds to the composition of the respective embeddings.

Suppose that for each $\alpha < \kappa$, $L_\alpha = (L_\alpha,\leq)$ is a complete lattice. 
Suppose that a family of projections $h^\alpha_\beta : L_\alpha \to L_\beta$ for 
$\beta < \alpha < \kappa$ is specified such that $h^\beta_\gamma \circ h^\alpha_\beta = h^\alpha_\gamma$,
for all $\gamma <  \beta < \alpha$. Then we say that the complete lattices $L_\alpha$, $\alpha < \kappa$,
form an \emph{inverse system}, c.f. \cite{Scott},\footnote{The complete lattices 
and projections of an inverse system
of \cite{Scott} are continuous, 
and the ordinal $\kappa$ is $\omega$, the least infinite ordinal. 
Inverse systems of complete lattices over arbitrary directed partial orders 
are considered in \cite{compendium}, where following \cite{Scott}, 
the projections are usually assumed to be continuous as well.} with projections $h^\alpha_\beta$,
$\beta < \alpha < \kappa$.  

For the rest of this section, suppose that we are given such an inverse system 
of complete lattices. We denote the embedding corresponding to each 
$h^\alpha_\beta$ by $k^\alpha_\beta$.  As noted above, it follows that 
$k^\alpha_\beta \circ k^\beta_\gamma = k^\alpha_\gamma$, 
for all $\gamma < \beta < \alpha < \kappa$. Also, for each $\beta < \alpha < \kappa$,
$h^\alpha_\beta$ preserves all infima and $k^\alpha_\beta$ preserves all suprema. 
We will sometimes also suppose that the projections $h^\alpha_\beta$ are completely additive,
or at least locally completely additive, see below. 
It will be convenient to define $h^\alpha_\alpha$ and $k^\alpha_\alpha$ 
for $\alpha < \kappa$ as the identity function $L_\alpha \to L_\alpha$.

Let $L_\infty$ be the \emph{inverse limit} determined by the above
inverse system.  Thus, $L_\infty \subseteq \prod_{\alpha < \kappa} L_\alpha$ is the 
collection of all $\kappa$-sequences $x = (x_\alpha)_{\alpha < \kappa}$ in $\prod_{\alpha < \kappa} L_\alpha$ 
with $h^\alpha_\beta(x_\alpha) = x_\beta$ for all $\beta < \alpha< \kappa$, ordered 
by the relation $\leq$ defined pointwise. A sequence in $L_\infty$ will be referred to as a 
`compatible sequence'. Since the functions $h^\alpha_\beta$ preserve all 
infima, $L_\infty$ is indeed a complete lattice in which the
infimum $\bigwedge X$ of any set $X \subseteq L_\infty$ is formed pointwise. This follows by noting
that the pointwise infimum of any set of compatible 
sequences is compatible, since the functions $h^\alpha_\beta$ preserve all 
infima.  The least element of $L_\infty$ 
is the compatible sequence $(\bot_\alpha)_{\alpha < \kappa}$
composed of the least elements of the lattices $L_\alpha$.
The greatest element is the sequence $(\top_\alpha)_{\alpha < \kappa}$,
where for each $\alpha < \kappa$, $\top_\alpha$ is the greatest element of $L_\alpha$. 
If the functions $h^\alpha_\beta$, $\beta < \alpha < \kappa$, are all completely additive, 
then the supremum $\bigvee X$ of any set $X$ of sequences in $L_\infty$ is 
also formed pointwise. 
To facilitate notation, we will denote the supremum and the infimum of a 
subset $X$ of $L_\alpha$ by $\bigvee_\alpha X$ and $\bigwedge_\alpha X$,
respectively.

For each $\alpha < \kappa$, let $h^\infty_\alpha$
denote the function $L_\infty \to L_\alpha$ mapping each $x \in L_\infty$
to the $\alpha$-component $x_\alpha$ of $x$. 
These functions form  a \emph{cone} over the inverse system 
$h^\alpha_\beta: L_\alpha \to L_\beta$, since $h^\alpha_\beta \circ h^\infty_\alpha 
= h^\infty_\beta$ for all $\beta < \alpha < \kappa$.

\begin{lem}
\label{lem-proj}
Suppose that the complete lattices $L_\alpha$, $\alpha < \kappa$, form an inverse 
system with projections $h^\alpha_\beta : L_\alpha \to L_\beta$, $\beta < \alpha < \kappa$,
and limit $L_\infty$. 
Then each function $h^\infty_\alpha: L_\infty \to L_\alpha$ for $\alpha < \kappa$ is also a 
projection.
\end{lem}

{\sl Proof.} 
For each $x\in L_\alpha$, where $\alpha < \kappa$, let $k^\infty_\alpha(x) = (y_\beta)_{\beta < \kappa}$ 
with $y_\beta = h^\alpha_\beta(x)$ if $\beta \leq \alpha$, 
and $y_\beta = k^\beta_\alpha(x)$ if $\beta > \alpha$, where 
$k^\beta_\alpha$ is the embedding corresponding to $h^\beta_\alpha$. 
Then $k^\infty_\alpha(x) \in L_\infty$ and clearly $h^\infty_\alpha(k^\infty_\alpha(x)) = x$. 
And if $z = (z_\beta)_{\beta < \kappa}$ is in $L_\infty$, 
then $k^\infty_\alpha(h^\infty_\alpha(z)) \leq z$, since 
if $\beta \leq \alpha$ then the $\beta$-component of 
$k^\infty_\alpha(h^\infty_\alpha(z))$ is $z_\beta$, 
and if $\beta > \alpha$, then the $\beta$-component of 
$k^\infty_\alpha(h^\infty_\alpha(z))$ is $k^\beta_\alpha(z_\alpha) \leq 
z_\beta$, since $z_\alpha = h^\beta_\alpha(z_\beta)$ and
$(h^\alpha_\beta, k^\alpha_\beta)$ is a projection-embedding pair.   
Thus, $h^\infty_\alpha:L_\infty \to L_\alpha$ is a projection
with corresponding embedding $k^\infty_\alpha: L_\alpha \to L_\infty$. 
\eop 

It follows that the functions $h^\infty_\alpha$ preserve all infima 
and the functions $k^\infty_\alpha$ preserve all suprema.

The complete lattice $L_\infty$ has the following property. 
Suppose that $L$ is a complete lattice and the 
functions $g_\alpha : L \to L_\alpha$ form another cone, where $\alpha < \kappa$,
so that $g_\beta = h^\alpha_\beta \circ g_\alpha$ for all $\beta < \alpha < \kappa$.  
Then there is a unique function $g: L\to L_\infty$ 
such that $h^\infty_\alpha \circ g = g_\alpha$, for all $\alpha < \kappa$. 
Indeed, for each $y \in L$, $g(y) = (g_\alpha(y))_{\alpha < \kappa}$. 
If the functions $g_\alpha$, $\alpha < \kappa$, are monotone,
then so is this \emph{mediating} function $g$, and vice versa.  
We will call the functions $h^\infty_\alpha$, $\alpha < \kappa$,
\emph{limit functions}, or \emph{limit projections}.

\begin{lem}
\label{lem-galois}
Suppose that the complete lattices $L_\alpha$, $\alpha < \kappa$, form an inverse 
system with projections $h^\alpha_\beta: L_\alpha \to L_\beta$, $\beta < \alpha < \kappa$,
and limit $L_\infty$. 
Let $L$ be a complete lattice with a cone of projections $g_\alpha: L \to L_\alpha$
and corresponding embeddings $f_\alpha : L_\alpha \to L$, for each $\alpha < \kappa$, 
and let $g$ denote the mediating function $L \to L_\infty$, $y \mapsto (g_\alpha(y))_{\alpha < \kappa}$. 
Define $f: L_\infty \to L$ by 
$f(x) = \bigwedge \{y : y \in L,\  \forall \gamma < \kappa\ x_\gamma \leq g_\gamma(y)\}
= \bigwedge \{y : y \in L,\ x \leq g(y)\}$ 
for all $x = (x_\gamma)_{\gamma< \kappa} \in L_\infty$. Then the pair of 
functions $g$ and $f$ forms a Galois connection between $L_\infty$ and $L$. 
\end{lem}

{\sl Proof.}
Indeed, we have already noted that $g$ is monotone, and it is clear that $f$ is 
also monotone. 
Let $x = (x_\gamma)_{\gamma < \kappa} \in L_\infty$. Then for all $\alpha < \kappa$, 
\begin{eqnarray*}
g_\alpha (f(x)) 
&=& 
g_\alpha (\bigwedge \{y: y \in L,\ \forall \gamma < \kappa\ x_\gamma \leq g_\gamma(y)\})\\
&=& 
\bigwedge_\alpha \{g_\alpha(y): y \in L,\ \forall \gamma < \kappa\ x_\gamma \leq g_\gamma(y)\},
\end{eqnarray*}
since $g_\alpha$ preserves arbitrary infima. It is clear that 
$x_\alpha \leq \bigwedge_\alpha \{g_\alpha(y) \in L: \forall \gamma < \kappa\ x_\gamma \leq g_\gamma(y)\}$, 
thus $x_\alpha \leq g_\alpha(f(x))$. 
Since this holds for all $\alpha < \kappa$, it follows that the identity function
over $L_\infty$ is less than or equal to $g \circ f$ with respect to the pointwise ordering. We still need to prove that 
$f \circ g$ is less than or equal to the identity 
function over $L$. But for all $y \in L$,
\begin{eqnarray*}
f(g(y)) 
&=& 
\bigwedge \{ z: z \in L,\ g(y) \leq g(z)\}\\
&\leq& y,
\end{eqnarray*}
since $g(y) \leq g(y)$. 
\eop 

\begin{remark}
\label{rem-galois}
For later use we note that if the mediating function $g$ of Lemma~\ref{lem-galois}
is surjective, or if for each $x = (x_\gamma)_{\gamma < \kappa}$ in $L_\infty$ and $\alpha < \kappa$ 
there is some $y \in L$ with $x_\alpha = g_\alpha(y)$ and $x_\gamma \leq g_\gamma(y)$
for all $\gamma < \kappa$, then $g$ is a projection. Indeed, if either of these 
assumptions applies, then $x_\alpha = g_\alpha(f(x))$ for all $\alpha < \kappa$
and $x \in L$, where $f$ is defined as in Lemma~\ref{lem-galois}.  
\end{remark}

If the projections $h^\alpha_\beta$, $\beta < \alpha < \kappa$, satisfy 
a weak form of complete additivity, then we can prove that the mediating morphism $g$ is 
in fact a projection. Call a monotone function $L' \to L$ \emph{locally completely additive}
if for all $Y \subseteq L'$ and $x \in L$ with $h(Y) = \{x\}$ (i.e., $Y$ is nonempty and 
$h$ maps each element of $Y$ to $x$), it holds that $h (\bigvee Y) = x$. It is clear that when 
a function $h: L' \to L$ is completely additive, then 
it is locally completely additive. 

\begin{expl}
{\rm 
Let $L$ be the $2$-element lattice $\{\bot,\top\}$ with $\bot < \top$,
and let $L'$ be the lattice of nonnegative integers, ordered as usual, 
endowed with a greatest element $\infty$, so that $L'$ is a
complete lattice. The (surjective) function $L' \to L$ that maps each 
nonnegative integer to $\bot$ and $\infty$ to $\top$ is a projection 
but not locally completely additive. 
}
\end{expl}

\begin{expl}
{\rm 
There exist finite and hence complete lattices $L$ and $L'$ 
with a locally (completely) additive projection $L' \to L$ 
which is not (completely) additive, i.e., does not preserve binary
suprema. Let $L'$ have 7 elements, the multisets $\emptyset, \{a\}, \{b\}, 
\{a,a\}, \{a,b\}, \{b,b\}$, ordered by inclusion, together with a 
greatest element $\top$. Let $L$ consist of the sets $\emptyset,\{a\},\{b\},\{a,b\}$,
ordered by inclusion, together with a greatest element $\top$. Let $h$ map 
$\{a,a\}$ to $\{a\}$, $\{b,b\}$ to $\{b\}$, and let $h$ be the identity function otherwise. 
Then $h$ is a locally completely additive projection that is not completely additive, 
since the supremum of $\{a,a\}$ and $\{b,b\}$ in $L'$ is $\top$, 
while the supremum of $\{a\}$ and $\{b\}$ in $L$ is $\{a,b\}$. 
}
\end{expl} 

\begin{lem}
Let $L$ and $L'$ be complete lattices and $g: L' \to L$ 
 monotone and surjective. Then $g$ is locally completely additive iff 
 $\bigvee g^{-1}(x) \in g^{-1}(x)$ for all $x \in L$.
\end{lem} 

{\sl Proof.} 
Suppose first that $g$ is locally completely additive. Let $x \in L$ 
and $Y = g^{-1}(x)$. Then $g(Y) = \{x\}$, thus $g(\bigvee Y) = x$ 
and $\bigvee g^{-1}(x) = \bigvee Y \in g^{-1}(x)$, 
since $g$ is locally completely additive.

Suppose now that $\bigvee g^{-1}(x) \in g^{-1}(x)$ 
for all $x \in L$. Let $x \in L$ and $Y \subseteq L'$
with $g(Y) = \{x\}$. Then $Y$ is not empty, say $y_0\in Y$. 
Since $y_0 \leq \bigvee Y \leq \bigvee g^{-1}(x)$
and $g$ is monotone, it holds that 
$$x = g(y_0) \leq g(\bigvee Y) \leq g(\bigvee g^{-1}(x)) = x.$$
Thus, $g(\bigvee Y) = x$.
\eop 

\begin{lem}
\label{lem-locally additive projections}
Let $L_\infty$ be the limit of the inverse system of complete lattices 
$L_\alpha$, $\alpha < \kappa$, with locally completely additive
projections $h^\alpha_\beta: L_\alpha \to L_\beta$, $\beta < \alpha < \kappa$. 
Then the limit projections $h^\infty_\beta: L_\infty \to L_\beta$, $\beta < \kappa$,
are also locally completely additive. 
\end{lem}

{\sl Proof.} Suppose that $x \in L_\beta$ and $Y = (h^\infty_\beta)^{-1}(x)$, 
where $\beta < \kappa$ is  a fixed ordinal. We need to prove that 
$h^\infty_\beta (\bigvee Y) = x$.

For each $\alpha$ with $\beta < \alpha < \kappa$, let $Y_\alpha = (h^\alpha_\beta)^{-1}(x)$.
If $\beta < \alpha < \alpha' < \kappa$, then $(h^{\alpha'}_\beta)^{-1}(x) 
= (h^{\alpha'}_\alpha)^{-1}((h^\alpha_\beta)^{-1}(x))$, hence   
 $Y_{\alpha'} = (h^{\alpha'}_\alpha)^{-1} (Y_\alpha)$. Moreover, $h^{\alpha'}_\alpha(Y_{\alpha'}) = Y_\alpha$. 
Also, $Y = (h^\infty_\alpha)^{-1}(Y_\alpha)$ and $h^\infty_\alpha(Y) = Y_\alpha$
for all $\alpha$ with $\beta < \alpha < \kappa$. 

For each $\alpha$ with $\beta < \alpha < \kappa$, define $y_\alpha = \bigvee Y_\alpha$.
When $\alpha \leq \beta$, let $y_\alpha = h^\beta_\alpha(x)$.  We intend to show that the sequence 
$(y_\alpha)_{\alpha < \kappa}$ is compatible, so that $y = (y_\alpha)_{\alpha < \kappa}$ is 
in $L_\infty$. 
 
We have $y_\alpha \in Y_\alpha$ for all $\alpha$ 
with $\beta < \alpha < \kappa$, since $h^\alpha_\beta$ is locally completely additive.
 Thus, if $\beta < \alpha <  \alpha'$,
then $h^{\alpha'}_\alpha(y_{\alpha'}) = y_\alpha$, since $h^{\alpha'}_\alpha (y_{\alpha'})$ 
is necessarily the greatest element of $Y_\alpha$. When $\alpha < \alpha' <\kappa$ 
with $\alpha \leq \beta$, then $h^{\alpha'}_\alpha(y_{\alpha'}) = h^\beta_\alpha(x_\beta) = y_\alpha$.
Thus, $y \in L_\infty$.

We claim that $y = \bigvee Y$ in $L_\infty$. We have already shown that $y \in L_\infty$.
We know that for each $\alpha$ with $\beta < \alpha < \kappa$, 
it holds that $y_\alpha = \bigvee Y_\alpha$. Thus, our claim holds
if for all such $\alpha$, $Y_\alpha$ is equal to the set of all $\alpha$-components 
of the sequences in $Y$. But this is clear, since $Y_\alpha = h^\infty_\alpha(Y)$. 

It follows now that $h^\infty_\beta$ is locally completely additive. \eop

\begin{lem}
\label{lem-locally completely additive mediating} 
Let $L_\infty$ be the limit of an inverse system of complete lattices 
$L_\alpha$, $\alpha < \kappa$, with locally completely additive
projections $h^\alpha_\beta: L_\alpha \to L_\beta$, $\beta < \alpha < \kappa$. 
Suppose that $L$ is a complete lattice and 
the locally completely additive projections $g_\alpha : L \to L_\alpha$, $\alpha < \kappa$, form a cone. 
Then the unique mediating function $g: L \to L_\infty$ is a projection. 
\end{lem} 

{\sl Proof.} 
We already know that $g$ is a projection if it is surjective, 
cf. Lemma~\ref{lem-galois} and Remark~\ref{rem-galois}. Below we prove that $g$ is 
indeed surjective. We will also give a new description of the corresponding 
embedding.

For each $\alpha < \kappa$, let $f_\alpha$ denote the embedding corresponding to $g_\alpha$.
When $x = (x_\alpha)_{\alpha < \kappa}$ is in $L_\infty$, define $f(x) = \bigvee_{\alpha < \kappa} f_\alpha(x_\alpha)$.
We prove that $g(f(x)) = x$ for all $x \in L_\infty$ and that $f$ is the embedding corresponding to $g$.

So let $x = (x_\alpha)_{\alpha < \kappa}$ in $L_\infty$. If $\beta < \alpha < \kappa$, then 
\begin{eqnarray*}
f_\beta(x_\beta) 
&=& 
\bigwedge \{y: y \in L,\ x_\beta \leq g_\beta(y)\}\\
&\leq &
\bigwedge \{y: y \in L,\ x_\alpha \leq g_\alpha(y)\}\\
&=&
f_\alpha(x_\alpha),
\end{eqnarray*}
since if $x_\alpha \leq g_\alpha(y)$ for some $y \in L$, 
then $x_\beta = h^\alpha_\beta(x_\alpha) \leq h^\alpha_\beta(g_\alpha(y)) = g_\beta(y)$. 
Hence the sequence $(f_\alpha(x_\alpha))_{\alpha < \kappa}$ is increasing. 
If $\gamma \leq \alpha < \kappa$,
then 
\begin{eqnarray*}g_\gamma(f_\alpha(x_\alpha))
&=& 
h^\alpha_\gamma (g_\alpha(f_\alpha(x_\alpha)))\\
&=&  
h^\alpha_\gamma(x_\alpha)\\
&=&  
x_\gamma.
\end{eqnarray*}
Thus, 
\begin{eqnarray*}
g_\gamma (\bigvee_{\alpha < \kappa} f_\alpha (x_\alpha)) 
&=&
g_\gamma(\bigvee_{\gamma \leq \alpha < \kappa} f_\alpha (x_\alpha))\\
&=& x_\gamma,
\end{eqnarray*}
since $g_\gamma$ is locally completely additive. 
Since this holds for all $\gamma < \kappa$, we conclude that $g(f(x)) = x$
for all $x \in L_\infty$.

Suppose now that $y \in L$. Then 
\begin{eqnarray*}
f(g(y)) &=&
f((g_\alpha(y))_{\alpha < \kappa}) \\
&=& \bigvee_{\alpha < \kappa} f_\alpha(g_\alpha(y))\\
&\leq & y,
\end{eqnarray*} 
since $f_\alpha(g_\alpha(y)) \leq y$ for all $\alpha < \kappa$. 
\eop 

\begin{cor}
Under the assumptions of the previous lemma, for all $(x_\alpha)_{\alpha < \kappa} \in L_\infty$, 
$$\bigwedge\{y : y \in L,\ \forall \alpha < \kappa\ x_\alpha \leq g_\alpha(y)\}
= \bigvee_{\alpha < \kappa} f_\alpha(x_\alpha).$$
\end{cor}

\begin{lem}
\label{lem-locally completely additive mediating2} 
Let $L_\infty$ be the limit of an inverse system of complete lattices 
$L_\alpha$, $\alpha < \kappa$, with locally completely additive
projections $h^\alpha_\beta: L_\alpha \to L_\beta$, $\beta < \alpha < \kappa$. 
Suppose that $L$ is a complete lattice and the locally completely additive projections
$g_\alpha : L \to L_\infty$, $\alpha < \kappa$, form a cone. 
Then the unique mediating function $g: L \to L_\infty$ is a 
locally completely additive projection. 
\end{lem} 

{\sl Proof.} Let $h^\infty_\alpha : L_\infty \to L_\alpha$, $\alpha < \kappa$, be the 
limit functions defined above. We know that they are locally completely additive projections. 
Suppose that $Y \subseteq L$, $x = (x_\alpha)_{\alpha < \kappa} \in L_\infty$ 
and $g(Y) = \{x\}$. Then $g_\alpha(Y) = h^\infty_\alpha(g(Y)) = x_\alpha$,
hence $g_\alpha(\bigvee Y) = x_\alpha$
for all $\alpha < \kappa$, since $g_\alpha$ is locally 
completely additive. Since this holds for all $\alpha$, we have $g(\bigvee Y) = x$. 
On the other hand, $g$ is a projection by Lemma~\ref{lem-locally completely additive mediating}. 
\eop

We now consider inverse systems with completely additive projections. 

\begin{lem}
\label{lem-completely additive projections}
Let $L_\infty$ be the limit of an inverse system of complete lattices 
$L_\alpha$, $\alpha < \kappa$, with projections $h^\alpha_\beta: L_\alpha \to L_\beta$, 
$\beta < \alpha < \kappa$. 
Suppose that each $h^\alpha_\beta$ is completely additive.
Then the limit projections $h^\infty_\alpha: L_\infty \to L_\alpha$, $\alpha < \kappa$, are also 
completely additive. 
\end{lem}

{\sl Proof.} Let $X \subseteq L_\infty$ and $\alpha < \kappa$. 
Let $X_\alpha$ denote the set of $\alpha$-components of the 
sequences in $X$.
Since the supremum of $X$ in $L_\infty$ is 
formed pointwise,  $h^\infty_\alpha(\bigvee X) = \bigvee_\alpha X_\alpha 
= \bigvee_\alpha h^\infty_\alpha(X)$.
\eop

\begin{lem}
\label{lem-completely additive mediating}
Let $L_\infty$ be the limit of an inverse system of complete lattices 
$L_\alpha$, $\alpha < \kappa$, with projections $h^\alpha_\beta: L_\alpha \to L_\beta$, $\beta < \alpha < \kappa$. 
Suppose that each $h^\alpha_\beta$ for $\beta < \alpha< \kappa$ 
is also completely additive. Let $L$ be a complete 
lattice and suppose that the completely additive 
functions $g_\alpha : L \to L_\alpha$, $\alpha < \kappa$ 
form a cone.  
Then the mediating function $g: L\to L_\infty$ is also completely 
additive. 
\end{lem} 

{\sl Proof.} Indeed, for all $X \subseteq L$, 
$g(\bigvee X) = (g_\alpha(\bigvee X))_{\alpha < \kappa} 
= (\bigvee_\alpha g_\alpha(X))_{\alpha < \kappa}  
= \bigvee\{(g_\alpha(x))_{\alpha < \kappa} : x \in X\} = 
\bigvee g(X)$. 
\eop

\begin{remark}
{\rm 
Let $L_\infty$ be the limit of an inverse system of complete lattices 
$L_\alpha$, $\alpha < \kappa$, with completely additive projections 
$h^\alpha_\beta: L_\alpha \to L_\beta$ having corresponding
embeddings $k^\alpha_\beta : L_\beta \to L_\alpha$, $\beta < \alpha < \kappa$. 
We know that the limit functions $h^\infty_\alpha : L_\infty \to L_\alpha$, 
$\alpha < \kappa$, are also completely additive projections. For 
each $\alpha < \kappa$, let $k^\infty_\alpha: L_\alpha \to L_\infty$ 
denote the embedding corresponding to $h^\infty_\alpha$. 
Then the complete lattices $L_\alpha$, $\alpha < \kappa$, equipped 
with the embeddings $k^\alpha_\beta : L_\beta \to L_\alpha$, $\beta < \alpha < \kappa$,
form a \emph{direct system}. Moreover, $L_\infty$, equipped with the  
embeddings $k^\infty_\alpha : L_\alpha \to L_\infty$ has the following universal 
property. Given a complete lattice $L$ together with a family of 
completely additive functions $f_\alpha : L_\alpha \to L$ for $\alpha < \kappa$ such that 
$f_\alpha \circ k^\alpha_\beta = f_\beta$ for all $\beta < \alpha < \kappa$, 
there is a unique completely additive function $f: L_\infty \to L$ with $f \circ k_\alpha^\infty   
= f_\alpha$ for all $\alpha < \kappa$. 
Indeed, given $x = (x_\alpha)_{\alpha < \kappa}$ in $L_\infty$, we have
$f(x) = \bigvee_{\alpha < \kappa} f_\alpha(x_\alpha)$. 
And if each $f_\alpha$ is an embedding, then so is $f$. 
See also \cite{Scott} and Theorem IV-5.5 in \cite{compendium}, where 
continuity is required instead of complete additivity, so that 
the mediating function $f$ is continuous. 
}
\end{remark}

\section{Inverse limit models}
\label{sec-inverse limit models} 

In this section, our aim is to prove that the limit of an inverse system of 
complete lattices with locally completely additive projections determines a model. Moreover,
when the projections of the inverse system are completely additive, then the
limit determines a strong model.

Suppose that $L_\alpha$, $\alpha < \kappa$, is an inverse system 
of complete lattices with projections 
$h^\alpha_\beta: L_\alpha \to L_\beta$, $\beta < \alpha < \kappa$.
Let $L_\infty$ denote the limit of the inverse system 
with limit projections $h^\infty_\alpha : L_\infty \to L_\alpha$.

For each $\alpha < \kappa$, define the relation $\sqsubseteq_\alpha$ 
on $L_\alpha$ by $x \sqsubseteq_\alpha y$ iff $x \leq y$ and 
$h^\alpha_\beta(x) = h^\alpha_\beta(y)$  for all $\beta < \alpha$. 
Clearly, $\sqsubseteq_\alpha$ is a partial ordering of $L_\alpha$ 
which is included in the complete lattice order $\leq$ on $L_\alpha$.

We also define preorderings $\sqsubseteq_\alpha$ on $L_\infty$.
For all $\alpha < \kappa$ and $x = (x_\gamma)_{\gamma < \kappa}$ 
and $y = (y_\gamma)_{\gamma < \kappa}$ in $L_\infty$, let 
$x \sqsubseteq_\alpha y$ iff $x_\alpha \sqsubseteq_\alpha y_\alpha$ in $L_\alpha$,
i.e., when $x_\alpha \leq y_\alpha$ and $x_\beta = y_\beta$ 
for all $\beta < \alpha$. Thus, for all $x,y \in L_\infty$ and 
$\alpha < \kappa$, if $x \sqsubseteq_\alpha y$ then $h^\infty_\alpha (x) \sqsubseteq_\alpha h^\infty_\alpha(y)$, 
hence $h^\infty_\alpha(x) \leq h^\infty_\alpha(y)$ and $h^\infty_\beta (x) = h^\infty_\beta(y)$ for all $\beta < \alpha$.

By the above definition, each $\sqsubseteq_\alpha$ is a preorder,
so that $L_\infty$ is a stratified complete lattice. 
Moreover, the intersection of all equivalence relations 
$=_\alpha$, determined by the preorderings $\sqsubseteq_\alpha$, $\alpha < \kappa$, 
is the identity relation on $L_\infty$. Thus, A1 and A2 hold.
We show that A3 holds.

\begin{lem}
\label{lem-A3}
Let $L_\infty$ be the stratified complete lattice determined by the limit of an inverse system 
of complete lattices $L_\alpha$, $\alpha < \kappa$, 
with projections
$h^\alpha_\beta: L_\alpha \to L_\beta$, $\beta < \alpha < \kappa$. 
Then for all $x \in L_\infty$ and $\alpha < \kappa$ there is some $y \in L_\infty$ 
with $x=_\alpha y$ and such that for all $z\in L_\infty$, if $x \sqsubseteq_\alpha z$ 
then $y \leq z$. 
\end{lem}

{\sl Proof.} 
Suppose that $x = (x_\gamma)_{\gamma < \kappa}$ 
is in $L_\infty$. Let $\alpha < \kappa$ and define $y = (y_\gamma)_{\gamma < \kappa}$ 
as follows. Let $y_\gamma = x_\gamma$ for all $\gamma \leq \alpha$. And if 
$\alpha < \gamma$, define $y_{\gamma} = k^\gamma_\alpha(x_\alpha)$,
where $k^\gamma_\alpha$ is the embedding determined by the projection $h^\gamma_\alpha$.
Note that $y \in L_\infty$ and $y =_\alpha x$, since $y_\alpha = x$.
 In fact, $y = k^\infty_\alpha(h^\infty_\alpha(x))$,
where the limit projection $h^\infty_\alpha$ and corresponding embedding $k^\infty_\alpha$ were defined above.

Let $z = (z_\gamma)_{\gamma < \kappa}$ in $L_\infty$. Suppose that 
$x \sqsubseteq_\alpha z$. Then $y_\alpha = x_\alpha \leq z_\alpha$ and 
$y_\beta = x_\beta = z_\beta$ for all $\beta < \alpha$.
Suppose now that $\alpha < \beta < \kappa$. Then $y_\beta = k^\beta_\alpha(y_\alpha)
= k^\beta_\alpha(x_\alpha) \leq k^\beta_\alpha(z_\alpha) \leq z_\beta$,
since $x_\alpha \leq z_\alpha$ and $k^\beta_\alpha$ is monotone, 
and since $h^\beta_\alpha(z_\beta) = z_\alpha$.
Thus, $y \leq z$ and $y\sqsubseteq_\alpha z$. 
\eop

Under the assumptions of Lemma~\ref{lem-A3}, we denote $x|_\alpha =
k^\infty_\alpha(h^\infty_\alpha(x))$ for all $x \in L_\infty$ and $\alpha < \kappa$. 

\begin{lem}
Let $L_\infty$ be the stratified complete lattice determined by the limit of an inverse system 
of complete lattices $L_\alpha$, $\alpha < \kappa$, 
with projections
$h^\alpha_\beta: L_\alpha \to L_\beta$, $\beta < \alpha < \kappa$. 
Then for all $x \in L_\infty$, it holds that $x = \bigvee_{\alpha < \kappa} x|_\alpha$.  
\end{lem} 

{\sl Proof.} For all $\alpha < \kappa$, $x|_\alpha \leq x$ and $x =_\alpha x|_\alpha$,
i.e., the $\alpha$-component of $x$ agrees with the $\alpha$-component of $x|_\alpha$. 
Thus, $\bigvee_{\alpha < \kappa}  x|_\alpha \leq x$ and $x \leq y$ whenever $x|_\alpha \leq y$
for all $\alpha < \kappa$. \eop 

It is also clear that A5 and A6 hold. We thus have:

\begin{cor}
\label{cor-xxx}
Let $L_\infty$ be the stratified complete lattice determined by the limit of an inverse system 
of complete lattices $L_\alpha$, $\alpha < \kappa$, 
with projections $h^\alpha_\beta: L_\alpha \to L_\beta$, $\beta < \alpha < \kappa$.
Then $L_\infty$, equipped with the relations $\sqsubseteq_\alpha$, $\alpha < \kappa$, 
satisfies A1, A2, A3, A5, A6. Moreover, $x = \bigvee_{\alpha < \kappa} x|_\alpha$
for all $x \in L_\infty$.  
\end{cor} 

\begin{lem}
\label{lem-A4}
Suppose that $L_\infty$ is the stratified complete lattice determined by the limit of an inverse system 
of complete lattices $L_\alpha$, $\alpha < \kappa$, 
with locally completely additive projections
$h^\alpha_\beta: L_\alpha \to L_\beta$, $\beta < \alpha < \kappa$.
Suppose that $X$ is a nonempty subset of $L_\infty$, $y \in L_\infty$ and $\alpha < \kappa$
with $X =_\alpha y$, i.e., $x =_\alpha y$ for all $x \in X$. Then $\bigvee X =_\alpha y$. 
\end{lem} 

{\sl Proof.} Since $X =_\alpha y$, it holds that 
$h^\infty_\alpha(X) = y$. Since by Lemma~\ref{lem-locally additive projections}, $h^\infty_\alpha$ 
is locally completely additive, we conclude that $h^\infty_\alpha(\bigvee X) = y$,
i.e., $\bigvee X =_\alpha y$. \eop

\begin{prop}
\label{prop-inverse}
Let $L_\alpha$, $\alpha < \kappa$, be the stratified complete lattice determined by 
an inverse system of complete lattices
with projections $h^\alpha_\beta : L_\alpha \to L_\beta$,
$\beta < \alpha < \kappa$. 
Then the inverse limit $L_\infty$ is a model satisfying the axioms A1--A6
iff each of the projections $h^\alpha_\beta$ for $\beta < \alpha < \kappa$ 
is locally completely additive. 
Moreover, in this case, the limit functions $h^\infty_\alpha : L_\infty \to L_\alpha$,
$\alpha < \kappa$, are locally completely additive projections.  
\end{prop} 

{\sl Proof.} 
Suppose first that the projections $h^\alpha_\beta$
are locally completely additive. Then $L$ is a model by 
Corollary~\ref{cor-xxx} and Lemma~\ref{lem-A4}.
Moreover, the limit functions $h^\infty_\alpha$ are locally completely
additive projections by Lemmas~\ref{lem-proj}
and \ref{lem-locally additive projections}. 

Suppose now that $L_\infty$ is a model. We want to prove that each $h^\alpha_\beta$ 
is locally completely additive. First we show that each $h^\infty_\alpha$ is.
Suppose that $Y \subseteq L_\infty$ is not empty and $h^\infty_\alpha(Y) = x$.
Then $Y =_\alpha k^\infty_\alpha (x)$, since the $\alpha$-component of 
each sequence in $Y$ is $x$ as is the $\alpha$-component of 
$k^\infty_\alpha(x)$.  Since $L_\infty$ is a model, it follows that 
$\bigvee Y =_\alpha k^\infty_\alpha(x)$. This means that the $\alpha$-component of 
$\bigvee Y$ agrees with the $\alpha$-component $x$ of $k^\infty_\alpha(x)$, 
hence $h^\infty_\alpha(\bigvee_\alpha Y) = x$. 

Suppose now that $\beta < \alpha < \kappa$ 
and $x\in L_\beta$. Let $Y = (h^\alpha_\beta)^{-1}(x)$ and 
$Z = (h^\infty_\beta)^{-1}(x) = (h^\infty_\alpha)^{-1}(Y)$.
Since $h^\infty_\beta$ is locally completely additive, 
$\bigvee Z \in Z$ and thus $h^\infty_\alpha(\bigvee Z) \in Y$. 
But $Y = h^\infty_\alpha(Z) \leq h^\infty_\alpha(\bigvee Z)$,
thus $\bigvee_\alpha Y = h^\infty_\alpha(\bigvee Z) \in Y$.
\eop

If the projections $h^\alpha_\beta$ are completely additive, then the stronger
version A4$^*$ of axiom A4 holds.

\begin{lem}
\label{lem-A4star}
Suppose that $L_\infty$ is the model determined by the limit of an inverse system 
of complete lattices $L_\alpha$, $\alpha < \kappa$, 
with completely additive projections
$h^\alpha_\beta: L_\alpha \to L_\beta$, $\beta < \alpha < \kappa$.
Suppose that $\alpha < \kappa$ and $x_i \sqsubseteq_\alpha y_i$ in $L_\infty$ 
for all $i \in I$. Then $\bigvee_{i \in I} x_i \sqsubseteq_\alpha \bigvee_{i \in I} y_i$.
\end{lem}

{\sl Proof.} By our assumption, the $\beta$-component of $x_i$ 
agrees with the $\beta$-component of $y_i$ for all $i\in I$ and $\beta < \alpha$.
Moreover, for all $i \in I$, 
the $\alpha$-component of $x_i$ is less than or equal to 
the $\alpha$-component of $y_i$. Since the supremum is 
formed pointwise (cf. Lemma~\ref{lem-completely additive projections}), 
it follows that for all $\beta < \alpha$,
the $\beta$-component of $\bigvee_{i \in I} x_i$ agrees 
with the $\beta$-component of $\bigvee_{i \in I} y_i$, 
and the $\alpha$-component of $\bigvee_{i \in I} x_i$ 
is less than or equal to the $\alpha$-component of 
$\bigvee_{i \in I} y_i$. 
Thus $\bigvee_{i \in I} x_i \sqsubseteq_\alpha \bigvee_{i \in I} y_i$.
\eop

\begin{prop}
\label{prop-inverse2}
Let $L_\alpha$, $\alpha < \kappa$, be an inverse system of complete lattices 
with projections $h^\alpha_\beta : L_\alpha \to L_\beta$,
$\beta < \alpha < \kappa$, and denote by $L_\infty$ the 
stratified complete lattice  determined by
limit of the system.  
If the projections $h^\alpha_\beta$ are 
completely additive, then the inverse limit $L_\infty$ is a strong model, i.e., it satisfies 
 A1, A2, A3, A4$^*$, A5 and A6.  
Moreover, the limit projections $h^\infty_\alpha : L_\infty \to L_\alpha$,
$\alpha < \kappa$, are completely additive.  

Conversely, if $L_\infty$ is a strong model, then the projections $h^\alpha_\beta$,
$\beta < \alpha< \kappa$, are completely additive. 
\end{prop} 

{\sl Proof.} Suppose that the projections $h^\alpha_\beta$, $\beta < \alpha < \kappa$, 
are completely additive. Then they are locally completely additive, hence $L_\infty$ is
a model by Proposition~\ref{prop-inverse}.  Thus, by Lemma~\ref{lem-A4star}, $L_\infty$ 
is a strong model. 

Suppose now that $L_\infty$ is a strong model. Let $X \subseteq L_\infty$ and $\alpha < \kappa$.
Since $x =_\alpha k^\infty_\alpha(h^\infty_\alpha(x))$ for all $x \in X$ and $L_\infty$ is a strong model,
we have $\bigvee X =_\alpha \bigvee k^\infty_\alpha(h^\infty_\alpha(X)) = 
k^\infty_\alpha(\bigvee_\alpha(h^\infty_\alpha(X))$, where the last equality holds since 
$k^\infty_\alpha$ preserves all suprema. Applying $h^\infty_\alpha$ to both sides this 
gives $h^\infty_\alpha(\bigvee X) = h^\infty_\alpha(k^\infty_\alpha(\bigvee_\alpha h^\infty_\alpha(X)))
= \bigvee_\alpha h^\infty_\alpha(X)$. Thus each $h^\infty_\alpha$ is completely additive. 
It follows that for each $\beta < \alpha < \kappa$, $h^\alpha_\beta = h^\infty_\beta \circ k^\infty_\alpha$,
$h^\alpha_\beta$ is also completely additive.
\eop

\begin{expl}\label{expl-V_omega}
{\rm 
Let $\kappa = \Omega$ be the least uncountable ordinal, and for each $\alpha < \Omega$, let $L_\alpha$ 
be the linearly ordered lattice $F_0 < \ldots < F_\alpha < 0 <  T_\alpha < \ldots < T_0$.
For all $\beta < \alpha < \Omega$, define $h^\alpha_\beta : L_\alpha \to L_\beta$ by 
$h^\alpha_\beta(F_\gamma) = F_\gamma$ and
$h^\alpha_\beta(T_\gamma) = T_\gamma$, for all $\gamma \leq \beta$, and let $h^\alpha_\beta(x) = 0$, otherwise. 
Then all of the assumptions of Proposition~\ref{prop-inverse2} are satisfied
so that $L_\infty$ is a strong model. In fact, 
$L_\infty$ is isomorphic to $V_\Omega$. An isomorphism $L_\infty \to V_\Omega$ 
is given by the assignment that maps the sequence 
$(0,0,\ldots ,F_\alpha, F_\alpha,\ldots )$ to $F_\alpha$, 
the sequence $(0,0,\ldots, T_\alpha,T_\alpha,\ldots)$ to $T_\alpha$, where $\alpha < \Omega$ 
and the first $F_\alpha$ 
or $T_\alpha$ occurs in position $\alpha$,
and the $0$-sequence $(0,0,\ldots)$ to $0$. 
}
\end{expl}

We will prove in Section~\ref{sec-rep} that every model satisfying the axioms A1--A6 is 
isomorphic to a model determined by the limit of an inverse system of complete
lattices with locally completely additive projections. Moreover, 
we will prove that every strong model is isomorphic
to a model determined by the limit of an inverse system of complete lattices 
with completely additive projections. 

\section{Some properties of models}
\label{sec-cons}

In this section, we establish several consequences of the axioms. These results 
will be used in our proof of the fact that every model is isomorphic
to an inverse limit model. Suppose that $L$ satisfies the axioms A1--A6.
For each $x\in L $ and $\alpha < \kappa$, let 
$[x]_\alpha = \{y \in L : x=_\alpha y\}$. 
Moreover, for each $\alpha < \kappa$, let $L|_\alpha = \{x|_\alpha : x \in L\}$.

\begin{lem}
\label{lem1}
For each $x\in L$ and $\alpha < \kappa$, it holds that $x =_\alpha x|_\alpha$, $x|_\alpha \leq x$,
and $x|_\alpha$ is the $\leq$-least element of $[x]_\alpha$.
\end{lem} 

{\sl Proof.} 
The first claim is clear, since by A3, $x=_\alpha x|_\alpha$. 
Suppose that $y \in [x]_\alpha$. Then $x =_\alpha y$ and so $x \sqsubseteq_\alpha y$. 
Thus, $x|_\alpha \leq y$, by A3 and the definition of $x|_\alpha$. 
In particular, since $x \in [x]_\alpha$, it holds that $x|_\alpha \leq x$. \eop

\begin{cor}
For all $x \in L$ and $\alpha < \kappa$, it holds that 
$x|_\alpha = \bigwedge [x]_\alpha = \bigwedge \{ y: x \sqsubseteq_\alpha y\}$. 
\end{cor}

\begin{cor}
\label{corxvee}
For all $x \in L$, $\bigvee_{\alpha < \kappa} x|_\alpha \leq x$. 
\end{cor}

\begin{cor}
\label{cor1}
For all $x,y\in L$ and $\alpha < \kappa$, it holds that
$x \sqsubseteq_\alpha y$ iff $x|_\alpha \sqsubseteq_\alpha y$ iff $x \sqsubseteq_\alpha y|_\alpha$ 
iff $x|_\alpha \sqsubseteq_\alpha y|_\alpha$.
\end{cor} 

{\sl Proof.} This follows from the fact $x =_\alpha x|_\alpha$ and $y=_\alpha y|_\alpha$,
proved in Lemma~\ref{lem1}. 
\eop 

\begin{cor}
\label{cor2}
For all $x,y\in L$ and $\alpha < \kappa$, it holds that 
$x =_\alpha y$ iff $x|_\alpha =_\alpha y$ iff $x|_\alpha =_\alpha y|_\alpha$.
Moreover, $x =_\alpha y$ iff $x|_\alpha = y|_\alpha$.
\end{cor}

{\sl Proof.} 
This follows from Corollary~\ref{cor1} and Lemma~\ref{lem1},
by noting that if $x =_\alpha y$, then $[x]_\alpha = [y]_\alpha$,
so $x|_\alpha$ and $y|_\alpha$ are $\leq$-least elements of the same set. 
\eop

\begin{lem}
\label{lem2}
Suppose that $x \in L$ and $\alpha < \beta < \kappa$. Then $x|_\alpha =_\alpha x|_\beta$ 
and $x|_\alpha \leq x|_\beta$.
\end{lem} 

{\sl Proof.} By Lemma~\ref{lem1}, it holds that $x|_\alpha =_\alpha x =_\beta x|_\beta$. Since 
by A1 the relation $=_\beta$ is included in the relation $=_\alpha$, we 
conclude that $x|_\alpha =_\alpha x|_\beta$. Since 
$[x]_\beta \subseteq [x]_\alpha$, the $\leq$-least element 
of $[x]_\alpha$ is less than or equal to the $\leq$-least 
element of $[x]_\beta$. 
Thus, by Lemma~\ref{lem1}, $x|_\alpha \leq x|_\beta$.
\eop

\begin{lem}
\label{lem3}
Suppose that $x\in L$ and $\alpha, \beta < \kappa$.
If $\alpha \leq \beta$ then $(x|_\alpha)|_\beta = x|_\alpha$.
If $\beta < \alpha$ then $(x|_\alpha)|_\beta = x|_\beta$.
\end{lem}

{\sl Proof.} By Lemma~\ref{lem1}, it holds that $(x|_\alpha)|_\beta \leq x|_\alpha$. 
If $\alpha \leq \beta$ then, since $x|_\alpha =_\alpha x$, 
by A1 we have $[x|_\alpha]_\beta \subseteq [x]_\alpha$,
hence the $\leq$-least element of $[x]_\alpha$ is less than or equal 
to the $\leq$-least element of $[x|_\alpha]_\beta$.
Thus, by Lemma~\ref{lem1}, $x|_\alpha \leq (x|_\alpha)|_\beta$. 
We conclude that $(x|_\alpha)|_\beta = x|_\alpha$.

Suppose now that $\beta < \alpha$. Then by $x|_\alpha =_\alpha x$,
which holds by Lemma~\ref{lem1}, and by the fact that the relation $=_\alpha$ 
is included in $=_\beta$, which holds by A1, 
we have $[x|_\alpha]_\beta = [x]_\beta$. Thus, $(x|_\alpha)|_\beta = x|_\beta$ by 
Lemma~\ref{lem1}.
\eop 

\begin{cor}
\label{corLalpha} 
For all $x\in L$ and $\alpha < \kappa$, $x \in L|_\alpha$ iff $x = x|_\alpha$.
\end{cor} 

{\sl Proof.} Recall that $L|_\alpha = \{y|_\alpha : y \in L\}$. Thus, if 
$x = y|_\alpha$ is in $L|_\alpha$, then $x|_\alpha = (y|_\alpha)|_\alpha = y|_\alpha = x$.
If $x = x|_\alpha$, then clearly $x\in L|_\alpha$. \eop

\begin{cor}
\label{cor=alpha}
For all $x,y \in L|_\alpha$, $x =_\alpha y$ iff $x = y$. 
\end{cor}

{\sl Proof.} Suppose that $x,y \in L|_\alpha$. 
Then $x = x |_\alpha$ and $y = y|_\alpha$.
We conclude by Corollary~\ref{cor2}. 
\eop

\begin{lem}
\label{lemsqsubseteqalpha}
For all $x,y \in L$ and $\alpha < \kappa$, if $x \sqsubseteq_\alpha y$ then $x|_\alpha \leq y|_\alpha$.
\end{lem} 

{\sl Proof.} If $x \sqsubseteq_\alpha y$ then by $y =_\alpha y|_\alpha$,
also $x \sqsubseteq_\alpha y|_\alpha$, hence $x|_\alpha \leq y|_\alpha$ by A3. \eop 

\begin{cor}
\label{corsqsubseteqalpha}
For all $\alpha < \kappa$ and $x,y \in L|_\alpha$, if $x \sqsubseteq_\alpha y$ then 
$x \leq y$. 
\end{cor}

The above facts were all consequences of the first and the third axiom. 
We will now make use of A2 and A4 in order to prove a strengthened 
version of Corollary~\ref{corxvee}.

\begin{lem}
\label{lem4}
For all $x \in L$ and $\alpha < \kappa$, $x = \bigvee_{\alpha < \kappa} x|_\alpha$.
\end{lem}

{\sl Proof.} 
Let $\gamma < \kappa$ be any ordinal.
By Lemma~\ref{lem2}, the sequence $(x|_\alpha)_{\alpha < \kappa}$ 
is an increasing chain in $L$.
Thus $\bigvee_{\alpha < \kappa} x|_\alpha = \bigvee_{\gamma \leq \alpha < \kappa}x|_\alpha$. 
But for all $\alpha$ with $\gamma \leq \alpha$, $x|_\alpha =_\gamma x|_\gamma =_\gamma x$ 
by Lemma~\ref{lem2} and Lemma~\ref{lem1}.
Hence, by A4, 
$\bigvee_{\gamma \leq \alpha < \kappa} x|_\alpha =_\gamma x$
and thus $\bigvee_{\alpha < \kappa} x|_\alpha =_\gamma x$. 
Since this holds for all $\gamma < \kappa$, we conclude by A2 that 
$x = \bigvee_{\alpha < \kappa} x|_\alpha$.
\eop

\begin{lem}
\label{lem-newlem}
For all $\alpha < \kappa$, nonempty families $x_i \in L$, $i \in I$,  and $y \in L$,
if $x_i|_\alpha = y$ for all $i \in I$, then $(\bigvee_{i \in I} x_i)|_\alpha = y$
\end{lem} 

{\sl Proof}. This is clear from A4 and Corollary~\ref{cor=alpha}, since our assumption 
implies that $y \in L|_\alpha$. \eop 

\begin{remark}
\label{rem-newrem}
{\rm 
A certain converse of Lemma~\ref{lem-newlem} also holds. If A1, A2, A3 and the condition formulated 
in Lemma~\ref{lem-newlem} hold, and if $y \in L$ and $x_i \in L$
with $x_i =_\alpha y$ for all $i \in I$, where $I$ is a nonempty set, 
then by Corollary~\ref{cor=alpha}, $x_i|_\alpha = y|_\alpha$ for all $i \in I$, hence $(\bigvee_{i \in I} x_i)|_\alpha = y|_\alpha$.
By Corollary~\ref{cor=alpha} this means that $\bigvee_{i \in I} x_i =_\alpha y$, i.e., A4 holds. 
}
\end{remark}

The next facts also use A5. 

\begin{cor}
\label{cor3}
For all $x,y \in L$, $x \leq y$ iff $x|_\alpha \leq y|_\alpha$ 
for all $\alpha < \kappa$. 
\end{cor}

{\sl Proof.} Suppose that $x|_\alpha \leq y|_\alpha$ for all $\alpha < \kappa$. 
Then by Lemma~\ref{lem4}, $x = \bigvee_{\alpha < \kappa}x|_\alpha \leq 
\bigvee_{\alpha < \kappa} y|_\alpha = y$. The reverse direction holds by A5. 
\eop 

\begin{cor}
\label{cor4}
For all $x,y \in L$ and $\alpha < \kappa$, $x|_\alpha \leq y$ 
iff $x|_\alpha \leq y|_\alpha$.
\end{cor}

{\sl Proof.} This follows from Corollary~\ref{cor3} 
using the fact that $(x|_\alpha)|_\alpha = x|_\alpha$,
proved in Lemma~\ref{lem3}. \eop

The next facts depend on A6.

\begin{lem}
\label{lem6}
The following conditions are equivalent 
for  all $x,y \in L$ and $\alpha < \kappa$. 
\begin{itemize}
\item $x \sqsubseteq_\alpha y$.
\item $x|_\alpha \leq y|_\alpha$ and $x =_\beta y$ for all $\beta < \alpha$.
\item $x|_\alpha \leq y$ and $x =_\beta y$ for all $\beta < \alpha$.
\end{itemize}
\end{lem}

{\sl Proof.} Suppose that $x \sqsubseteq_\alpha y$. Then $x =_\beta y$ for all $\beta < \alpha$
by A1, and $x|_\alpha \leq  y|_\alpha$ by Lemma~\ref{lemsqsubseteqalpha}.
But if $x|_\alpha \leq y|_\alpha$, then also $x|_\alpha \leq y$, 
since by Lemma~\ref{lem1}, $y|_\alpha \leq y$.

Suppose that $x|_\alpha \leq y$ and $x =_\beta y$ for all $\beta < \alpha$. 
Then $x|_\alpha =_\beta y$ for all $\beta < \alpha$, 
 hence $x|_\alpha \sqsubseteq_\alpha y$ by A6. Thus, by  Lemma~\ref{lem1}, 
$x \sqsubseteq_\alpha y$.
\eop

\begin{cor}
\label{cor5}
For all $x,y \in L$ and $\alpha < \kappa$, $x|_\alpha  \sqsubseteq_\alpha y|_\alpha$ 
iff $x|_\alpha \leq y|_\alpha$ and $x|_\beta =_\beta y|_\beta$ for all $\beta < \alpha$. 
\end{cor}

{\sl Proof.} Immediate from Lemma~\ref{lem6} and Corollaries~\ref{corLalpha} and \ref{cor=alpha}.
\eop 

\begin{cor}
\label{cor6}
For all $x,y \in L|_\alpha$, $x \sqsubseteq_\alpha y$ iff $x|_\alpha \leq y|_\alpha$ 
and $x|_\beta = y|_\beta$ for all $\beta < \alpha$. 
\end{cor} 

{\sl Proof.} This is immediate from Corollary~\ref{cor5} and 
Corollary~\ref{corsqsubseteqalpha}. \eop

For each set
 $X \subseteq L$ and ordinal $\alpha < \kappa$, let us 
define $X|_\alpha = \{x|_\alpha : x \in X\}$. Note that this notation is 
consistent with the notation $L|_\alpha$ introduced earlier. 

Suppose now that $L$ is a strong model satisfying A4$^*$.
\begin{lem}
\label{lem5}
For all $X \subseteq L$ and $\alpha < \kappa$, 
$\bigvee X|_\alpha = (\bigvee X)|_\alpha$.
\end{lem}

{\sl Proof.} Let $X \subseteq L$ and $\alpha < \kappa$. Since by Lemma~\ref{lem1} $x =_\alpha x|_\alpha$ for 
all $x \in X$, it holds by A4$^*$ that $\bigvee X =_\alpha \bigvee X|_\alpha$.
Thus, $(\bigvee X)|_\alpha \leq \bigvee X|_\alpha$, again by Lemma~\ref{lem1}. 

Since $x \leq \bigvee X$ for all $x \in X$, by A5 we 
have $x|_\alpha \leq (\bigvee X)|_\alpha$ for all $x \in X$.
It follows that $\bigvee X|_\alpha \leq (\bigvee X)|_\alpha$.
\eop

\begin{remark}
\label{rem-compl}
{\rm 
Suppose that A1, A2 and A3 hold. Moreover, suppose that the 
property described in Lemma~\ref{lem5} holds. Then we can show 
that A4$^*$ and A5 hold. Thus, in the definition of strong 
models, these two axioms may be replaced by 
the property in Lemma~\ref{lem5}.

Indeed, if $x \leq y$ then for all $\alpha < \kappa$, 
$y|_\alpha = (x \vee y)|_\alpha = x|_\alpha \vee y|_\alpha$, hence 
$x|_\alpha \leq y|_\alpha$. And if $x_i =_\alpha y_i$ for all $i \in I$,
where $\alpha < \kappa$, then by Corollary~\ref{cor2}, $x_i|_\alpha = y_i|_\alpha$ for 
all $i \in I$, thus 
$(\bigvee_{i \in I} x_i)|_\alpha = \bigvee_{i \in I}  x_i|_\alpha = \bigvee_{i \in I} y_i|_\alpha = 
(\bigvee_{i \in I} y_i)|_\alpha$. We conclude that 
$\bigvee_{i \in I}  x_i =_\alpha \bigvee_{i \in I}  y_i$. 
}
\end{remark}

\section{An alternative axiomatization}
\label{sec-alternative}

We used axiom A3 to equip a model $L$ with an operation $|_\alpha : L \to L$ 
for each $\alpha < \kappa$, mapping $x\in L$ to $x|_\alpha$ in $L|_\alpha \subseteq L$. 
In this section we give an alternative axiomatization
using these operations $|_\alpha$ instead of the preorderings $\sqsubseteq_\alpha$.

\begin{thm}
\label{thm-alternative}
Suppose that $L$ is a model satisfying the axioms A1--A6. For each $\alpha < \kappa$ 
and $x\in L$, let $x|_\alpha$ be defined by the following property (cf. A3):
\begin{itemize}
\item C. $x|_\alpha =_\alpha x$ and for all $y\in L$, if $x\sqsubseteq_\alpha y$ 
then $x|_\alpha \leq y$. 
\end{itemize}
Then, equipped with the operations $|_\alpha : L \to L$ 
for $\alpha < \kappa$, the following hold:
\begin{itemize}
\item B1. For all $x \in L$ and $\beta \leq \alpha < \kappa$, $(x|_\alpha)|_\beta = x|_\beta$.
\item B2. For all $x,y \in L$ and $\alpha < \kappa$, if $x \leq y$ then $x|_\alpha \leq y|_\alpha$.
\item B3. For all $x \in L$, $x = \bigvee_{\alpha < \kappa} x|_\alpha$. 
\item B4. For all $\alpha < \kappa$ and $y$ and $x_i\in L,\ i \in I$, 
where $I$ is a nonempty index set, 
if $x_i|_\alpha = y$ then $(\bigvee_{i \in I} x_i)|_\alpha = y$. 
\end{itemize}
Moreover, the following holds: 
\begin{itemize}
\item D. For each $\alpha < \kappa$ and $x,y \in L$, it holds that $x \sqsubseteq_\alpha y$ 
iff $x|_\alpha \leq y|_\alpha$ and $x|_\beta = y|_\beta$ for all $\beta < \alpha$. 
\end{itemize}

Suppose that $(L,\leq)$ is a complete lattice equipped with 
a family of functions $|_\alpha : L \to L$, $\alpha < \kappa$,  satisfying the axioms B1--B4. 
For each $\alpha < \kappa$, define the relation $\sqsubseteq_\alpha$ on $L$ by 
the condition D. 
Then, equipped with these relations $\sqsubseteq_\alpha$, $L$ is a model satisfying the axioms A1--A6.
Moreover, C holds.  
\end{thm}

{\sl Proof.} 
We have already proved that when $L$ is a model satisfying the axioms A1--A6, 
then equipped with the operations $|_\alpha: L \to L$, $\alpha < \kappa$, 
uniquely defined by C, $L$ satisfies B1--B4. In fact, B2 is the same as A5.
 Moreover, D holds. (See Lemma~\ref{lem3}, 
Lemma~\ref{lem4}, Lemma~\ref{lem-newlem} and Corollary~\ref{cor6}.)

Suppose now that $L$ is a complete lattice equipped with a family of functions $|_\alpha : L \to L$,
$\alpha < \kappa$, satisfying B1--B4. Define the relations $\sqsubseteq_\alpha$,
$\alpha < \kappa$, by D. Then each of the relations $\sqsubseteq_\alpha$, $\alpha < \kappa$, 
is clearly a preordering, and if $\beta < \alpha$, then $\sqsubseteq_\alpha$ is contained in $=_\beta$. 
Thus A1 holds. 

In order to prove that A2 holds, note first that  if $x \leq y$ then $x|_\alpha \leq y|_\alpha$
for all $\alpha < \kappa$, by B2, and if $x|_\alpha \leq y|_\alpha$  for all $\alpha < \kappa$,
then $x \leq y$, by B3. Thus, 
$x \leq y$ iff $x|_\alpha \leq y|_\alpha$ for all $\alpha < \kappa$, and 
$x = y$ iff $x|_\alpha = y|_\alpha$ for all $\alpha < \kappa$
iff $x =_\alpha y$ for all $\alpha < \kappa$, 
proving A2. 

Now we prove A3. First note that for all $\alpha < \kappa$ and $x \in L$, $x =_\alpha x|_\alpha$,
since by B1, $(x|_\alpha)|_\beta = x|_\beta$ for all $\beta \leq \alpha$. Moreover, if $x \sqsubseteq_\alpha y$, 
then by D and B3, $x|_\alpha \leq y|_\alpha \leq y$.

Axiom A4 holds by B4 and Remark~\ref{rem-newrem}. A5 holds since it is the same as B2. 
Finally, axiom A6 holds, since if $x \leq y$ in $L$ and 
$x|_\beta = y|_\beta$ for all $\beta < \alpha$, where $\alpha < \kappa$, then, by B2,
 also $x|_\alpha \leq y|_\alpha$ and thus $x \sqsubseteq_\alpha y$
by D. \eop 

\begin{cor}
Suppose that $L$ is a strong model satisfying the axioms A1, A2, A3, A4$^*$, A5 and A6. For each $\alpha < \kappa$ 
and $x\in L$, let $x|_\alpha$ be defined by the property C above. 
Then, equipped with the operations $|_\alpha : L \to L$ 
for $\alpha < \kappa$, B1, B3 and the following hold: 
\begin{itemize}
\item B2$^*$. For all $X \subseteq L$ and $\alpha < \kappa$, $(\bigvee X)|_\alpha = \bigvee X|_\alpha$.
\end{itemize}
Moreover, D holds.

Suppose that $L$ is a complete lattice equipped with 
a family of functions $|_\alpha : L \to L$, $\alpha < \kappa$,  satisfying the axioms B1, B2$^*$ and B3. 
For each $\alpha < \kappa$, define the relation $\sqsubseteq_\alpha$ on $L$ by 
the condition D. 
Then, equipped with these relations $\sqsubseteq_\alpha$, $L$ is a strong model.
Moreover, C holds.  
\end{cor}

{\sl Proof.} One uses Lemma~\ref{lem5} and Remark~\ref{rem-compl}. \eop 

\begin{remark}
{\rm 
The proof of Theorem~\ref{thm-alternative} entails also the following result.

Suppose that $L$ is a stratified complete lattice satisfying the axioms A1, A2, A3, A5, A6 and B3,
where for each  $\alpha < \kappa$ 
and $x\in L$, $x|_\alpha$ is defined by the property C. 
Then, equipped with the operations $|_\alpha : L \to L$ 
for $\alpha < \kappa$, B1, B2 and D hold.

Suppose that $(L,\leq)$ is a complete lattice equipped with 
a family of functions $|_\alpha : L \to L$, $\alpha < \kappa$,  satisfying the axioms B1, B2, B3.  
For each $\alpha < \kappa$, define the relation $\sqsubseteq_\alpha$ on $L$ by 
the condition D. 
Then, equipped with these relations $\sqsubseteq_\alpha$, $L$ satisfies A1, A2, A3, A5 and A6. 
Moreover, C holds.  
}
\end{remark}

\section{The representation theorem} 
\label{sec-rep}

In this section, we prove that every model satisfying the axioms 
A1--A6 introduced in Section~\ref{sec1} is isomorphic to an inverse limit model. 
In our argument, we will make use of the properties of 
models established in the previous sections.

\begin{prop}
\label{prop1} Suppose that $L$ is a model satisfying A1--A6.
Then for each $\alpha < \kappa$, $L|_\alpha$, equipped with the 
ordering inherited from $L$, is a complete 
lattice. Moreover, for all $X \subseteq L|_\alpha$, the 
infimum $\bigwedge_\alpha X$ of $X$ in $L|_\alpha$ is $(\bigwedge X)|_\alpha$,
where $\bigwedge X$ is the infimum of $X$ in $L$. Similarly, the 
supremum $\bigvee_\alpha X$ of $X$ in $L|_\alpha$ is 
$(\bigvee X)|_\alpha$, where $\bigvee X$ is the supremum of $X$ 
in $L$. 
\end{prop}

{\sl Proof.} 
Suppose that $L$ is a model. Let $\alpha < \kappa$
and $X \subseteq L|_\alpha$.

Since by Lemma~\ref{lem1} (or B3),  $(\bigwedge X)|_\alpha \leq \bigwedge X$, 
we have $(\bigwedge X)|_\alpha  \leq X$. Suppose that $z \in L|_\alpha$ 
with $z \leq X$. 
Then $z \leq \bigwedge X$, hence $z \leq (\bigwedge X)|_\alpha$ 
by Corollary~\ref{corLalpha} and Corollary~\ref{cor4}, 
or B1 and B2. We have completed the proof of the fact 
that $(\bigwedge X)|_\alpha$ is the infimum of $X$ in $L|_\alpha$, 
i.e., $\bigwedge_\alpha X = (\bigwedge X)|_\alpha$. 

The proof of $\bigvee_\alpha X = (\bigvee X)|_\alpha$ is similar.
First, $X \leq \bigvee X$, hence $X \leq (\bigvee X)|_\alpha$
by Corollary~\ref{cor4}, or B1 and B2. And if $z \in L|_\alpha$ with $X \leq z$, 
then $\bigvee X  \leq z$, hence $(\bigvee X)|_\alpha \leq z$ by Lemma~\ref{lem1} (or B3).  
\eop 

\begin{expl}
\label{expl-rep1}
{\rm 
Suppose that $L$ is the 5-element lattice on the set $\{\bot,0,1,2,\top\}$, ordered so that $\bot$ and 
$\top$ are the least and the greatest element, respectively, moreover, $0 < 1$ and $0 < 2$, but 
there is no order relation between $1$ and $2$. Let $\sqsubseteq_0$ 
be the least preordering such that $\bot =_0 0$ holds. 
Let $\sqsubseteq_1$ be the least preordering such that $\bot \sqsubseteq_1 1$, 
and when $2 \leq \alpha < \kappa$, let $\sqsubseteq_\alpha$ be the identity relation. Then $L|_0$ is the sublattice 
of $L$ determined by the set $\{\bot,1,2,\top\}$, and for each $\alpha$ with $1 \leq \alpha < \kappa$, 
$L|_\alpha = L$. 
The function $|_0 : L\to L|_0$ maps $\bot$ and $0$ to $\bot$ and is the identity function 
otherwise. For each $\alpha$ with $1 \leq \alpha < \kappa$, $|_\alpha$ is the identity function $L\to L$. 
Then $L$ is a strong model. 
Note that it is not true that for all $x,y\in L|_0$, $x \wedge y = x \wedge_0 y$,
since $1 \wedge 2 = 0$ while $1 \wedge_0 2 = \bot$. However, $(1 \wedge 2)|_0 = 0|_0 = \bot$. 
}
\end{expl}

\begin{prop}
\label{prop2}
Suppose that $L$ is a model satisfying A1--A6.
For any ordinals $\alpha,\beta$ with $\beta \leq \alpha < \kappa$, 
define $h^\alpha_\beta : L|_\alpha \to L|_\beta$
by $h^\alpha_\beta(x) = x|_\beta$ for all $x\in L|_\alpha$. 
Then each of the functions $h^\alpha_\beta$ for $\beta \leq \alpha < \kappa$ is surjective. 
For all $\alpha < \kappa$, $h^\alpha_\alpha$ is the identity function $L|_\alpha \to L|_\alpha$,
and for all $\gamma < \beta < \alpha < \kappa$, 
$h^\beta_\gamma \circ h^\alpha_\beta = h^\alpha_\gamma$. Moreover, the following hold: 
\begin{itemize}
 \item For all $\beta < \alpha < \kappa$, $h^\alpha_\beta : L|_\alpha \to L|_\beta$ is a projection. 
\item 
  For all $\beta < \alpha < \kappa$, $h^\alpha_\beta$ is locally completely additive. 
\item For all $\alpha < \kappa$ and $x,y \in L|_\alpha$,
      $x \sqsubseteq_\alpha y$  iff $x \leq y$ and $h^\alpha_\beta(x) = h^\alpha_\beta(y)$
      for all $\beta < \alpha$. 
\end{itemize}
\end{prop}

{\sl Proof.} Suppose that $\beta \leq \alpha < \kappa$. For all $x \in L$, it holds by 
Lemma~\ref{lem3} (or B1) that $(x|_\alpha)|_\beta = x|_\beta$. Thus, $h^\alpha_\beta$ is 
surjective. 

By Lemma~\ref{lem3} (or B1), $h^\alpha_\alpha$ is the identity function $L|_\alpha \to L|_\alpha$
for all $\alpha < \kappa$. The fact that $h^\beta_\gamma \circ h^\alpha_\beta = h^\alpha_\gamma$ 
for all $\gamma < \beta < \alpha$ also follows from Lemma~\ref{lem3} (or B1), since for all
$x \in L$, $((x|_\alpha)|_\beta)|_\gamma = x|_\gamma = (x|_\alpha)|_\gamma$.

Suppose that $\beta < \alpha < \kappa$. If $x\leq y$ in $L|_\alpha$, then 
$x|_\beta \leq y|_\beta$ by A5 or B2. Thus, $h^\alpha_\beta$ is monotone. 
It follows from Lemma~\ref{lem3} that for all $\beta < \alpha < \kappa$, 
$L|_\beta \subseteq L|_\alpha$. Let $x \in L|_\beta$ and $y \in L|_\alpha$ with 
$x \leq y|_\beta$. Since $x \in L|_\beta$, it holds that $x = x|_\beta$, by 
Lemma~\ref{lem3} or B1. But again by Lemma~\ref{lem1} (or B3), 
$x|_\beta = x \leq y|_\beta \leq y$, so $x \leq y$. Also, if $x \leq y$, then $x \leq y|_\beta$.
Thus, $h^\alpha_\beta$ is a projection with corresponding embedding 
$k^\alpha_\beta : L|_\beta \to L|_\alpha$ being the inclusion function. 

Next we prove that each function $h^\alpha_\beta$ for $\beta < \alpha < \kappa$ 
is locally completely additive. To this end, suppose that 
$Y \subseteq L|_\alpha$ and $x \in L|_\beta$ with $h^\alpha_\beta(Y) = \{x\}$,
so that $Y$ is not empty and $y|_\beta = x$ for all $y \in Y$. Then, by Corollary~\ref{corLalpha}
and Corollary~\ref{cor=alpha}, or B1 and D, $y =_\beta x$ for all $y \in Y$,
i.e., $Y =_\beta x$. We conclude by A4 that $\bigvee Y =_\beta x$ and thus 
$(\bigvee Y)|_\beta = x$, again by Corollary~\ref{corLalpha}
and Corollary~\ref{cor=alpha}, or B1 and D. Thus, $h^\alpha_\beta(\bigvee_\alpha Y) = 
(\bigvee_\alpha Y)|_\beta = ((\bigvee Y)|_\alpha)|_\beta = (\bigvee Y)|_\beta = x$,
by Proposition~\ref{prop1} and either Lemma~\ref{lem3} or B1.

The last claim holds by Corollary~\ref{cor6} or D. \eop

We are now ready to prove the Representation Theorem, Theorem~\ref{thm-rep}. By Proposition~\ref{prop2}, 
for every model $L$ satisfying the axioms A1--A6, the 
complete lattices $L|_\alpha$ equipped with the locally completely additive 
projections $h^\alpha_\beta: L|_\alpha \to L|_\beta$ defined 
by $h^\alpha_\beta(x) = x|_\beta$ for all $x\in L|_\alpha$ and $\beta < \alpha < \kappa$ 
form an inverse system.
We can thus form the limit model $L_\infty$ as in Section~\ref{sec-inverse limit models}. 
We know that $L_\infty$ is a model satisfying the axioms A1--A6. 
But actually $L_\infty$ is isomorphic to $L$. 
   
\begin{thm}
\label{thm-rep}
Every model $L$ satisfying the axioms A1--A6 is isomorphic to the model determined by the limit of the 
inverse system of the complete lattices $L|_\alpha$, $\alpha < \kappa$, with locally 
completely additive projections  $h^\alpha_\beta: L|_\alpha \to L|_\beta$, defined 
by $h^\alpha_\beta(x) = x|_\beta$ for all $x\in L|_\alpha$,
 where $\beta <  \alpha < \kappa$. 
\end{thm}

{\sl Proof.}
Let $L_\infty$ denote the inverse limit.  We intend to show that 
$L$ is isomorphic to $L_\infty$. 
Recall that for each $\alpha < \kappa$, the limit projection $h^\infty_\alpha : L_\infty \to L|_\alpha$
maps a sequence $x \in L_\infty$ to its $\alpha$-component $x_\alpha$. 
We know from Proposition~\ref{prop-inverse} that these functions are locally completely additive projections 
and constitute a cone over the inverse system $h^\alpha_\beta: L|_\alpha \to L|_\beta$.

We define another cone. For each $\alpha < \kappa$, 
let  $f_\alpha : L \to L|_\alpha$ be defined by $f_\alpha (x) = x|_\alpha$. 
Note that each $f_\alpha$ is monotone and locally completely additive (Lemma~\ref{lem-newlem}) 
and a projection (Corollary~\ref{corLalpha} and Corollary~\ref{cor4}). 
Moreover, by Lemma~\ref{lem3} (or B1), 
$h^\alpha_\beta (f_\alpha (x)) = (x|_\alpha)|_\beta = x|_\beta = f_\beta (x)$
for all $\beta < \alpha$ and $x \in L$.
Thus, there is a unique function $f: L \to L_\infty$ with 
$h^\infty_\alpha \circ f = f_\alpha$ for all $\alpha < \kappa$.
We know that the function $f$, given by $f(x) = (x|_\alpha)_{\alpha < \kappa}$, is 
a locally completely additive projection (Lemma~\ref{lem-locally completely additive mediating} 
and Lemma~\ref{lem-locally completely additive mediating2}). 
By Corollary~\ref{cor3}, $f$ is an isomorphism.

To complete the proof, we still need to show that $f$ creates an isomorphism 
between $(L,\sqsubseteq_\alpha)$ 
and $(L_\infty,\sqsubseteq_\alpha)$ for each $\alpha$. But this is clear,
since for all $x,y \in L$, $x \sqsubseteq_\alpha y$ iff $x|_\alpha \sqsubseteq_\alpha y|_\alpha$,
as shown above (Corollary~\ref{cor1}). 
\eop 

\begin{expl}
\label{expl-rep2}
{\rm 
Let $L_0$ be the 4-element lattice 
that is not a chain, and when $0 < \alpha < \kappa$, let $L_\alpha$ be the 5-element lattice 
that is not a chain and has a unique minimal element greater than 
the least element. For each $0 < \alpha < \kappa$, let $h^\alpha_0 : L_\alpha \to L_0$ be the unique surjective monotone 
function that collapses the least element of $L_\alpha$ with the minimal element 
greater than the least element, and when $0 < \beta < \alpha < \kappa$, let 
$h^\alpha_\beta: L_\alpha \to L_\beta$ 
be the identity function. The functions $h^\alpha_\beta$, $\beta < \alpha < \kappa$, form a cone  
of projections preserving all suprema. The inverse limit $L_\infty$ is isomorphic to 
the lattice $L$ of Example~\ref{expl-rep1} and determines the same model. 
}
\end{expl}

\begin{expl}
\label{expl-per3}
{\rm 
Consider the model $L = V_\Omega$ defined above and recall Example~\ref{expl-V_omega}. 
Then for each $\alpha < \Omega$, $L|_\alpha$ is isomorphic to $L_\alpha$ and the functions
$h^\alpha_\beta : L|_\alpha \to L|_\beta$ given by $x \mapsto x|_\beta$ correspond to the functions 
$h^\alpha_\beta: L_\alpha \to L_\beta$ described in Example~\ref{expl-V_omega}.
} 
\end{expl}

\begin{cor}
\label{cor-rep}
Every strong model $L$ is is isomorphic to the model determined by the limit of the 
inverse system of the complete lattices $L|_\alpha$, $\alpha < \kappa$, with 
completely additive projections  $h^\alpha_\beta: L|_\alpha \to L|_\beta$, defined 
by $h^\alpha_\beta(x) = x|_\beta$ for all $x\in L|_\alpha$,
 where $\beta <  \alpha < \kappa$. 
\end{cor} 

{\sl Proof.} Let $L$ be a strong model. By Theorem~\ref{thm-rep}, $L$ is isomorphic
to the limit of the inverse system of the complete lattices $L|_\alpha$, 
$\alpha < \kappa$, with projections  $h^\alpha_\beta: L|_\alpha \to L|_\beta$
given above. Since $L$ is a strong model, the functions $h^\alpha_\beta$ 
are completely additive, cf. Proposition~\ref{prop-inverse2}. 
\eop

\begin{cor}
Let $L$ be a stratified complete lattice equipped with a preordering $\sqsubseteq_\alpha$ 
for each $\alpha < \kappa$. Then $L$ is a model satisfying the axioms A1--A6 
iff $L$ is isomorphic to the model determined by the limit of an inverse system of 
complete lattices $L_\alpha$, $\alpha < \kappa$, with locally completely 
additive projections $h^\alpha_\beta : L_\alpha \to L_\beta$, $\beta < \alpha < \kappa$. 
\end{cor}

\begin{cor}
Let $L$ be a stratified complete lattice equipped with a preordering $\sqsubseteq_\alpha$ 
for each $\alpha < \kappa$. Then $L$ is a strong model 
iff $L$ is isomorphic to the model determined by the limit of an inverse system of 
complete lattices $L_\alpha$, $\alpha < \kappa$, with completely 
additive projections $h^\alpha_\beta : L_\alpha \to L_\beta$, $\beta < \alpha < \kappa$. 
\end{cor}

\section{Some further properties of models}
\label{sec-further}

In this section, we establish several further properties of models. 
Some of these properties have been axioms 
in \cite{ERfp,ERwollic}, see Propositions~\ref{prop-p1}, \ref{prop-p3}. 
Some others, such as the ones formulated in Corollary~\ref{cor-lattice} and
Corollary~\ref{cor-leastfp}, were proved in \cite{ERfp} for a larger class 
of models. Our aim  here is to use the Representation Theorem to provide 
alternative proofs of these results. In Corollary~\ref{cor-lattice}, 
we will prove that if $L$ is a model, then it may naturally be equipped 
with another complete partial order $\sqsubseteq$. Then,
in Corollary~\ref{cor-leastfp}, we will show that certain weakly monotone functions 
over $L$ have least pre-fixed points with respect to the ordering $\sqsubseteq$, 
and that these least pre-fixed points are in fact fixed points.
Actually we will derive these facts 
from a new technical result formulated in Theorem~\ref{thm-general}, which 
also implies that the collection of all fixed points is in fact 
a complete lattice in itself w.r.t. the ordering $\sqsubseteq$,
cf. Corollary~\ref{cor-latticefp}.

In this section, we will without loss of generality 
suppose that a model $L$ is given as the model determined by the limit $L_\infty$ 
of an inverse system of complete lattices $L_\alpha$, $\alpha < \kappa$, with 
locally completely additive projections $h^\alpha_\beta : L_\alpha \to L_\beta$ 
and corresponding embeddings $k^\alpha_\beta : L_\beta \to L_\alpha$,
$\beta < \alpha < \kappa$. 
As before, we will denote the limit projection $L \to L_\alpha$ for $\alpha < \kappa$ by $h^\infty_\alpha$.
As noted above, the embeddings $k^\alpha_\beta$, as well as the embeddings $k^\infty_\alpha : L_\alpha \to L$,
corresponding to the projections $h^\infty_\alpha$, are locally completely additive. 
Recall that an element of an inverse limit model $L_\infty$ is a sequence 
$x = (x_\alpha)_{\alpha < \kappa}$,
which is compatible in the sense that $h^\alpha_\beta(x_\alpha) = x_\beta$ 
for all $\beta < \alpha < \kappa$. 
As opposed to previous sections, instead of $\bigvee_\alpha X$ and $\bigwedge_\alpha X$,
we will simply denote the supremum and infimum of a set $X \subseteq L_\alpha$, 
$\alpha < \kappa$, by $\bigvee X$ and $\bigwedge X$, respectively. 

The properties established in all models by Proposition~\ref{prop-p1} 
and Proposition~\ref{prop-p3} below have been axioms in \cite{ERfp}. 
We include these propositions in order to connect this paper with \cite{ERfp}. 

\begin{prop}
\label{prop-p1}
Suppose that $L$ is model satisfying A1--A6. Let $x \in L$,
$\alpha < \kappa$  and 
$X \subseteq (x]_\alpha = \{z : \forall \beta < \alpha\ x =_\beta z\}$.
Then there exists some $y \in (x]_\alpha$ with the following properties: 
\begin{itemize}
\item $X \sqsubseteq_\alpha y$ (i.e., $x \sqsubseteq_\alpha y$ for all $x \in X$),
\item For all $z\in (x]_\alpha$, if $X \sqsubseteq_\alpha z$ then $y \leq z$ and $y \sqsubseteq_\alpha z$.
\end{itemize}
\end{prop} 

{\sl Proof.} 
Before giving the proof, let us remark that for the notion of model as used
in this paper, Proposition~\ref{prop-p1} greatly simplifies. Using the above 
assumption and notation, since $X \subseteq (x]_\alpha$ and $y,z\in (x]_\alpha$, 
$X \sqsubseteq_\alpha y$ holds iff $X|_\alpha \leq y$, and similarly for 
$X \sqsubseteq_\alpha z$, moreover, $y \sqsubseteq_\alpha z$ iff $y|_\alpha \leq z$. 
See Lemma \ref{lem6}. But since $y|_\alpha \leq y$ (cf. Lemma~\ref{lem1}), we have $y \sqsubseteq_\alpha z$ 
and $y \leq z$ iff $y \leq z$. 
Thus, the above property amounts 
to the following assertion: for each $X \subseteq (x]_\alpha$ in a model $L$ satisfying 
A1--A6, there is some  $y\in (x]_\alpha$ with $X|_\alpha \leq y$ and such that for all 
$z\in L$, if $X|_\alpha \leq z$ then $y \leq z$.

In our proof, we make use of Theorem~\ref{thm-rep}. So without loss 
of generality suppose that $L= L_\infty$ is the model determined by the limit of an appropriate inverse 
system as described above. Then $x = (x_\beta)_{\beta < \kappa}$ is a compatible 
sequence, and $(x]_\alpha = \{(z_\beta)_{\beta < \kappa} \in L : 
\forall \beta < \alpha\ x_\beta = z_\beta \}$.

If $X$ is empty, let $y = \bigvee_{\gamma< \alpha} k^\infty_\gamma(x_\gamma)$, which is the 
least element of $(x]_\alpha$. 
Indeed, for any $\beta < \alpha$, $h^\infty_\beta(y) = h^\infty_\beta(\bigvee_{\gamma < \alpha} k^\infty_\gamma(x_\gamma)) = h^\infty_\beta (\bigvee_{\beta \leq \gamma < \alpha} k^\infty_\gamma(x_\gamma))$, since the sequence 
$(k^\infty_\gamma (x_\gamma))_{\gamma < \alpha}$ is increasing.  
But for all $\gamma$ with $\beta \leq \gamma < \alpha$, 
$h^\infty_\beta(k^\infty_\gamma(x_\gamma)) = h^\gamma_\beta(x_\gamma) = x_\beta$.
Thus, since $h^\infty_\beta$ is locally completely additive, we have
$h^\infty_\beta(\bigvee_{\gamma < \alpha} k^\infty_\gamma(x_\gamma))= \bigvee_{\beta \leq \gamma < \alpha} h^\infty_\beta(k^\infty_\gamma (x_\gamma)) 
= \bigvee_{\beta \leq \gamma < \alpha} x_\beta = x_\beta$.
And if $z = (z_\beta)_{\beta< \kappa} \in (x]_\alpha$, then $x_\beta = z_\beta = h^\infty_\beta(z)$ 
for all $\beta < \alpha$, hence $k^\infty_\beta (x_\beta) \leq z$ for all $\beta < \alpha$, 
so that $y = \bigvee_{\beta < \alpha} k^\infty_\beta(x_\beta) \leq z$.

If $X$ is not empty, then define $y = k^\infty_\alpha(\bigvee X_\alpha) = \bigvee k^\infty_\alpha (X_\alpha)$,
where $X_\alpha$ is the set of all $\alpha$-components of the elements of $X$. 
Since $(h^\infty_\alpha,k^\infty_\alpha)$ is a projection-embedding pair, $y$ is the least element of $L$ 
with $X_\alpha \leq h^\infty_\alpha(y)$,
or equivalently,  $\bigvee X_\alpha \leq  h^\infty_\alpha(y)$. To complete the proof, we 
still need to show that $y \in (x]_\alpha$. But for all $\beta < \alpha$, 
$h^\infty_\beta(y) = h^\infty_\beta(k^\infty_\alpha(\bigvee X_\alpha)) = 
h^\alpha_\beta(\bigvee X_\alpha) = x_\beta$, 
since $h^\alpha_\beta(X_\alpha) = x_\beta$ and $h^\alpha_\beta$ is locally completely additive. 
\eop

We will denote the element $y$ constructed above by $\bigsqcup_\alpha X$. 
Note that when $X$ is empty, $\bigsqcup_\alpha X$ depends on $x$, but 
if $X$ is not empty, then $\bigsqcup_\alpha X$ is independent of $x$. 
In particular, 
we may use the notation $\bigsqcup_\alpha X$ without specifying the element $x$ 
whenever $X$ is not empty and $z=_\beta z'$ holds for all $z,z'\in X$ and $\beta < \alpha$. 

We note that a short description of $\bigsqcup_\alpha X$ is $\bigvee (X|_\alpha \cup \{\overline{x}\})$,
where $\overline{x}$ is the least element of $(x]_\alpha$.

\begin{prop}
\label{prop-p3}
Suppose that $L$ is a strong model. Let $I$ be an arbitrary nonempty index 
set and $x_{i,n} \in L$ for all $i \in I$ and $n \geq 0$. Suppose that $\alpha < \kappa$ and 
$x_{i,n} \sqsubseteq_\alpha x_{i,n+1}$ for all $i \in I$ and $n \geq 0$.
Then $\bigvee_{i \in I} \bigsqcup_\alpha \{ x_{i,n} : n \geq 0 \}
=_\alpha \bigsqcup_\alpha \{ \bigvee_{i \in I} x_{i,n} : n \geq 0\}$.
\end{prop}  

{\sl Proof.} First note that $\bigsqcup_\alpha \{ \bigvee_{i \in I} x_{i,n} : n \geq 0\}$
exists, since by Proposition~\ref{lem-A4star}, $\bigvee_{i \in I}x_{i,n} \sqsubseteq_\alpha 
\bigvee_{i \in I}x_{i,n+1}$ for all $n\geq 0$, hence 
$\bigvee_{i\in I} x_{i,n} =_\beta 
\bigvee_{i \in I} x_{i,n+1}$ for all $n \geq 0$ and $\beta < \alpha$.

Again, we assume that $L$ is an inverse limit model. 
A routine calculation shows that both sides of the required equality 
are equal to $\bigvee_{i \in I, n \geq 0} (x_{i,n})_\alpha$, where 
for each $i\in I$ and  $n \geq 0$, 
$(x_{i,n})_\alpha$ is the $\alpha$-component of $x_{i,n}$. \eop 

\begin{remark}
{\rm 
Actually the above fact extends to all nonempty chains. 
Let $I$ be an arbitrary nonempty index and let $(J,\leq)$ be a nonempty chain.
Let $L$ be a model and  $x_{i,j} \in L$ for all $i \in I$ and $j \in J$. 
Suppose that $\alpha < \kappa$ and 
$x_{i,j} \sqsubseteq_\alpha x_{i,k}$ for all $i \in I$ and $j \leq k$ in $J$.
Then $\bigvee_{i \in I} \bigsqcup_\alpha \{ x_{i,j} : j \in J \}
=_\alpha \bigsqcup_\alpha \{ \bigvee_{i \in I} x_{i,j} : j \in J\}$.
}
\end{remark}

Suppose that $L$ is model satisfying A1--A6.. Following \cite{ERfp}, we define the relation $\sqsubseteq$ on $L$ 
by $x \sqsubseteq y$ iff $x = y$, or there is some $\alpha< \kappa$
with $x \sqsubset_\alpha y$, i.e., $x \sqsubseteq_\alpha y$
but $y \not \sqsubseteq_\alpha x$. When $L$ is an inverse limit model
and $x = (x_\alpha)_{\alpha < \kappa}$,  $y = (y_\alpha)_{\alpha < \kappa}$, 
this gives $x \sqsubseteq y$ iff either $x = y$, i.e., $x_\alpha = y_\alpha$ for all $\alpha < \kappa$,
or there is some $\alpha < \kappa$ with $x_\alpha < y_\alpha$ and $x_\beta = y_\beta$
for all $\beta < \alpha$.

\begin{lem}
\label{lem-sgsubseteq}
For every model $L$ satisfying A1--A6, the relation $\sqsubseteq$ is a 
partial order. Moreover, for every $x,y\in L$,
if $x \leq y$ then $x \sqsubseteq y$.  
\end{lem}

{\sl Proof.} Let $L$ be the model determined by the limit of an inverse system $L_\alpha$, $\alpha < \kappa$,
of complete lattices with locally completely additive projections 
$h^\alpha_\beta : L_\alpha \to L_\beta$, $\beta < \alpha< \kappa$. 
Let $x = (x_\alpha)_{\alpha < \kappa}$ 
and $y = (y_\alpha)_{\alpha < \kappa}$ in $L$. If $x = y$ then clearly $x \sqsubseteq y$.
Suppose that $x < y$. Then there is some $\alpha$ with $x_\alpha < y_\alpha$ and
$x_\beta = y_\beta$ for all $\beta < \alpha$. Thus, $x \sqsubset_\alpha y$ and $x \sqsubset y$.

It is clear $\sqsubseteq$ is reflexive and transitive. To prove that it is anti-symmetric, 
let $x,y$  in $L$.  Suppose that $x \sqsubseteq y$ and $y \sqsubseteq x$. If $x \neq y$
then there exist $\alpha,\beta < \kappa$
such that $x \sqsubset_\alpha y$ and $y \sqsubset_\beta x$. Then $x =_\gamma y$ 
for all $\gamma < \max\{\alpha,\beta\}$, which implies that $\alpha = \beta$
and hence $x_\alpha < y_\alpha$ and $y_\alpha < x_\alpha$, a contradiction.
Thus $x = y$. We note that when each $L_\alpha$ is linearly ordered, then 
$\sqsubseteq$ is a linear ordering of $L$. 
\eop 

Note that on inverse limit models, $\sqsubseteq$ is the lexicographic order.

\begin{expl}
{\rm 
Let $(L_0,\leq)$ be the 4-element lattice $\bot,0,1,\top$, ordered so that $\bot$ 
and $\top$ are the least and the greatest elements and $0$ and $1$ are incomparable
with respect to the ordering $\leq$. When $0 < \alpha < \kappa$, let $(L_\alpha,\leq)$ be the complete lattice 
whose set of elements is $\{\bot,0,1,\ldots, a,b,\top\}$, where $\bot$ and $\top$ 
are again the least and the greatest elements, respectively, moreover, the integers $0,1,\ldots$
form a chain with supremum $a$. The element $b$ is incomparable with $a$ and any integer element
with respect to $\leq$.

For each $\alpha$ with $0 < \alpha < \kappa$, let $h^\alpha_0$ map the element $\bot$ 
and all integer elements of $L_\alpha$ to $\bot$, $a$ to $0$, $b$ to $1$ and $\top$ to $\top$. 
When $0 < \beta < \alpha < \kappa$, let $h^\alpha_\beta$ 
be the identity function. The functions $h^\alpha_\beta$, $\beta  < \alpha < \kappa$ are 
projections, but the functions $h^\alpha_0$ are not locally completely 
additive. The lexicographic ordering of the inverse limit is not a lattice order,
since the elements $(0,a,a,\ldots)$ and $(1,b,b,\ldots)$ do not have 
an infimum. Indeed,
the lower bounds of these two sequences with respect to 
the lexicographic ordering $\sqsubseteq$ are those of the form 
$(\bot,n,n,\ldots)$, where $n$ is a nonnegative integer or $\bot$, and there is no greatest lower bound. 
}
\end{expl}

Below we will often make use of the following observation. Let  
 $L$ be the model determined by the limit of an inverse system of 
complete lattices $L_\alpha$, $\alpha < \kappa$, with 
locally completely additive projections $h^\alpha_\beta: L_\alpha \to L_\beta$, 
where $\beta < \alpha < \kappa$. Suppose that $\alpha < \kappa$ and $(x_\beta)_{\beta < \alpha}$ is 
a (partial) compatible sequence, so that $h^\beta_\gamma(x_\beta) = x_\gamma$ 
for all $\gamma < \beta < \alpha$. Then there is a least element $x_\alpha$
of $L_\alpha$ such that the sequence $(x_\beta)_{\beta \leq \alpha}$ 
is still compatible, namely $x_\alpha = \bigvee_{\beta < \alpha}k^\alpha_\beta(x_\beta)$.
Moreover, the set of all elements $x_\alpha$ with this property 
is a complete sublattice of $L_\alpha$ which is a closed interval.
Indeed, if $Y$ is a nonempty set of such elements of $L_\alpha$, then 
so is $\bigvee Y$, since $h^\alpha_\beta(Y) = \{x_\beta\}$ and thus $h^\alpha_\beta(\bigvee Y) =
\bigvee h^\alpha_\beta(Y) = x_\beta$ for all $\beta < \alpha$.
Finally, if $x_\alpha$ and $x'_\alpha$ in $L_\alpha$ satisfy $h^\alpha_\beta(x_\alpha) = h^\alpha_\beta(x'_\alpha)
= x_\beta$ for all $\beta < \alpha$, and if $x_\alpha \leq y \leq x'_\alpha$,
then by $h^\alpha_\beta(x_\alpha) \leq h^\alpha_\beta(y) \leq h^\alpha_\beta(x'_\alpha)$
we must have $h^\alpha_\beta(y) = x_\beta$ for all $\beta < \alpha$.

Suppose that $f: L \to L$, where $L$ is a model. Following \cite{ERfp}, we say that 
$f$ is \emph{$\alpha$-monotone} for some $\alpha < \kappa$
if $x \sqsubseteq_\alpha y$ implies $f(x) \sqsubseteq_\alpha f(y)$ for all 
$x,y\in L$. When $L$ is an inverse limit model as above, this means that 
if $x,y\in L$ are such that for each $\beta < \alpha$, the $\beta$-component of $x$ 
agrees with the corresponding component of $y$ and the $\alpha$-component 
of $x$ is less than or equal to the corresponding component of $y$, 
then the same hold for $f(x)$ and $f(y)$. 
Call a function $g: L_\alpha \to L_\alpha$ \emph{conditionally monotone} if 
for all $x,y\in L_\alpha$, if $h^\alpha_\beta(x) = h^\alpha_\beta(y)$ for all $\beta < \alpha$ 
and $x \leq y$, then $g(x) \leq g(y)$. 

\begin{lem}
Suppose that $L$ is a model determined by an inverse system of complete lattices $L_\alpha$,
$\alpha < \kappa$, with locally completely additive projections $h^\alpha_\beta: L_\alpha \to L_\beta$,
$\beta < \alpha < \kappa$. Let $f: L \to L$. Then $f$ is $\alpha$-monotone for all $\alpha < \kappa$ 
iff there exist conditionally monotone functions $f_\alpha: L_\alpha \to L_\alpha$, $\alpha < \kappa$, 
such that $f((x_\alpha)_{\alpha < \kappa} ) = (f_\alpha(x_\alpha))_{\alpha < \kappa}$ 
for all $(x_\alpha)_{\alpha < \kappa}$ in $L$.
\end{lem}

{\sl Proof}. In order to prove the sufficiency part of the lemma, suppose that $f: L \to L$ and 
$f_\gamma$, $\gamma < \kappa$,
is a family of conditionally monotone functions such that $f(x) = (f_\gamma(x_\gamma))_{\gamma < \kappa}$ 
for all $x = (x_\gamma)_{\gamma < \kappa} \in L$.  Let $\alpha < \kappa$ and $x,y \in L$
with $x \sqsubseteq_\alpha y $. Suppose that $x = (x_\gamma)_{\gamma < \kappa}$
and $y = (y_\gamma)_{\gamma < \kappa}$. We want to prove that $f(x) = x' \sqsubseteq_\alpha y ' = f(y)$. 
But for all $\beta < \alpha$, the $\beta$-component $x'_\beta$ of $x'$ agrees with the $\beta$-component 
$y'_\beta$ of $y'$, since by $x_\beta = y_\beta$ we have $x'_\beta = f_\beta(x_\beta) = f_\beta(y_\beta) = y'_\beta$.
Also, since $x_\alpha \leq y_\alpha$ and $f_\alpha$ is conditionally monotone, for the $\alpha$-components
we have $x'_\alpha = f_\alpha(x_\alpha) \leq f_\alpha(y_\alpha) = y'_\alpha$.

In order to prove the necessity part of the lemma, suppose that $f$ is $\alpha$-monotone for all 
$\alpha < \kappa$. For each $\alpha < \kappa$, define $f_\alpha: L_\alpha \to L_\alpha$ 
as the function $h^\infty_\alpha \circ f \circ k^\infty_\alpha$. If $x \leq y$ in $L_\alpha$
with $h^\alpha_\beta(x) = h^\alpha_\beta(y)$ for all $\beta < \alpha$,  
then for all $\beta < \alpha$, the $\beta$-component of 
 $k^\infty_\alpha(x)$ agrees with the $\beta$-component of $k^\infty_\alpha(y)$, 
 while the $\alpha$-component of $k^\infty_\alpha(x)$ is $x$ and the 
$\alpha$-component of $k^\infty_\alpha(y)$ is $y$, so that the $\alpha$-component of $k^\infty_\alpha(x)$
is less than or equal to the $\alpha$-component of $k^\infty_\alpha(y)$. Since $f$ 
is $\alpha$-monotone, the same holds for $f(k^\infty_\alpha(x))$ and $f(k^\infty_\alpha(y))$. 
In particular, the $\alpha$-component of $f(k^\infty_\alpha(x))$ is less than or equal to the 
$\alpha$-component of $f(k^\infty_\alpha(y))$, i.e., $f_\alpha(x) 
= h^\infty_\alpha(f(k^\infty_\alpha(x))) \leq h^\infty_\alpha(f(k^\infty_\alpha(y))) = f_\alpha(y)$.

We still need to prove that $f(x) = (f_\alpha(x_\alpha))_{\alpha < \kappa}$ 
for all $x = (x_\alpha)_{\alpha< \kappa}$ in $L$. Let $\alpha < \kappa$ be a fixed ordinal. 
Since $f$ is $\alpha$-monotone and $x =_\alpha k^\infty_\alpha(x_\alpha)$,
also $f(x) =_\alpha f(k^\infty_\alpha(x))$, hence 
 the $\alpha$-component of $f(x)$ agrees with the 
$\alpha$-component of $f(k^\infty_\alpha(x_\alpha))$, which is in turn equal to 
$f_\alpha(x_\alpha)$. Since $\alpha$ was an arbitrary ordinal less than $\kappa$,
this proves the required equality.
\eop 
 
In particular, when $f$ is $\alpha$-monotone for all $\alpha < \kappa$,
then $f_0$ is a monotone function over $L_0$.

A function $L \to L$ which is 
$\alpha$-monotone for all $\alpha < \kappa$ need not be monotone w.r.t. 
the partial order $\sqsubseteq$, cf. \cite {ERfp}.

\begin{remark}
{\rm 
Thus, if $L$ is an inverse limit model as above and $f: L \to L$ 
is $\alpha$-monotone for all $\alpha < \kappa$, then $f$ determines 
and is determined by a necessarily unique family of conditionally monotone functions 
$f_\alpha: L_\alpha \to L_\alpha$, $\alpha < \kappa$. Moreover, this family of functions is compatible
in the sense that $h^\alpha _\beta \circ f_\alpha = f_\beta \circ h^\alpha_\beta$ 
for all $\beta < \alpha < \kappa$. 

Conversely, if $f_\alpha$, $\alpha < \kappa$, is a compatible sequence of 
conditionally monotone functions, then for each compatible sequence $x = (x_\alpha)_{\alpha < \kappa}$,
the sequence $(f_\alpha(x_\alpha))_{\alpha < \kappa}$ is also compatible, and the 
function $f : L \to L$ defined by $f(x) = (f_\alpha(x_\alpha))_{\alpha < \kappa}$
for all $x = (x_\alpha)_{\alpha < \kappa}$ in $L$ is $\alpha$-monotone for all $\alpha < \kappa$.
} 
\end{remark}

We will also use the following fact. Suppose that $L$ is an inverse limit model 
as above and $f: L \to L$
is $\alpha$-monotone for all $\alpha < \kappa$. Suppose that $(x_\beta)_{\beta < \alpha}$ 
is a compatible sequence, so that $h^\beta_\gamma(x_\beta) = x_\gamma$
for all $\gamma < \beta <\alpha$. Consider the sublattice $Z_\alpha$ of $L_\alpha$ 
of those elements $x_\alpha$ such that the sequence $(x_\beta)_{\beta \leq \alpha}$
is still compatible. If for each $\beta < \alpha$, $x_\beta$ is a fixed point of $f_\beta$,
see below, then $f_\alpha$ maps $Z_\alpha$ into itself and is monotone on $Z_\alpha$.

Recall that a \emph{pre-fixed point} (resp. \emph{post-fixed point}) of a function $f$ over a partially ordered set 
$P$ is an element $x\in P$ with $f(x) \leq x$ (resp. $x \leq f(x)$). Moreover, $x$
is a \emph{fixed point} of $f$ if $f(x) = x$, i.e., when $x$ is both a pre-fixed point and a post-fixed point. 
By the well-known Knaster-Tarski fixed point theorem \cite{Daveyetal,Tarski}, every monotone endofunction
over a complete lattice has a least fixed point  which is also the least 
pre-fixed point. Dually, every monotone endofunction
over a complete lattice has a greatest fixed point, which is also the greatest
post-fixed point. And if $L$ is a complete lattice and $f: L \to L$
is monotone, then the fixed points of $f$ form a complete lattice. This 
immediately follows from the existence of the least fixed point using the 
fact that if $x$ is a post-fixed point, then there is a least pre-fixed point 
over $x$ which is a fixed point. More generally, if $X$ is a set of post-fixed points, then 
there is a least pre-fixed point over $X$ which is a fixed point. Of course, the dual 
statement also holds. 

In order to prove the above claim, suppose that $L$ is a complete lattice, $f: L \to L$ is 
monotone, and $X$ is a set of post-fixed points of $f$. Let 
$Z = \{ z \in L : X \leq z,\ f(z) \leq z\}$ and $y = \bigwedge Z$. 
We need to prove that $y$ is a fixed point of $f$. 

We have $X \leq y$ and thus $f(X) \leq f(y)$, hence $X \leq f(y)$ since $X$ 
is a set of post fixed points. And if $z \in Z$ then $y \leq z$, hence 
$f(y) \leq f(z) \leq z$. Since this holds for all $z \in Z$ 
and $y = \bigwedge Z$, we conclude thet $f(y) \leq y$. But then $f(y) \in Z$ 
and thus $y \leq f(y)$, proving $f(y) = y$.

\begin{thm}
\label{thm-general}
Let $L$ be a model satisfying the axioms A1--A6 and $f: L \to L$ be $\alpha$-monotone
for all $\alpha < \kappa$. Suppose that $X \subseteq L$ is a set of post-fixed 
points of $f$ with resect to the ordering $\leq$. Then there is a (necessarily unique) $y \in L$ 
with the following properties:
\begin{itemize}
\item $X \sqsubseteq y$ and $f(y) = y$,
\item for all $z\in L$, if $X \sqsubseteq z$ and $f(z) \sqsubseteq z$, then 
      $y \sqsubseteq z$.
\end{itemize} 
\end{thm} 
 
{\sl Proof.} Without loss of generality we may assume that $L$ is 
the model determined by the limit of an inverse system of complete
lattices $L_\alpha$, $\alpha < \kappa$, with locally completely additive
projections $h^\alpha_\beta: L_\alpha \to L_\beta$, $\beta < \alpha < \kappa$. 
Since $f$ is $\alpha$-monotone for all $\alpha < \kappa$, 
it is determined by a family of conditionally monotone functions $f_\alpha : L_\alpha \to L_\alpha$,
$\alpha < \kappa$. 

For each $\alpha < \kappa$, let $X_\alpha$ denote the set of all 
$\alpha$-components $x_\alpha$ of the elements $x$ of $X$. Define
\begin{eqnarray*}
Y_\alpha&=& \{z \in X_\alpha : \forall \beta < \alpha \ h^\alpha_\beta(z) = y_\beta\}
\end{eqnarray*}
and let $y_\alpha$ be the least (pre-)fixed point of $f_\alpha$ over $Y_\alpha$ in $Z_\alpha$, 
where $Z_\alpha$ is the set of all elements $z$ of $L_\alpha$ with $h^\alpha_\beta(z) = y_\beta$
for all $\beta < \alpha$. In particular, $Y_0 = X_0$ and $y_0$ is the least (pre-)fixed point of 
$f_0$ in $Z_0 = L_0$. 

It is clear that the sequence $y = (y_\alpha)_{\alpha < \kappa}$ is in $L$.
Moreover, $f(y) = y$, as each $y_\alpha$ is a fixed point of $f_\alpha$. The fact 
that $X \sqsubseteq y$ follows from the following: 

{\em Claim.} For all $x \in X$ and $\alpha < \kappa$, either $x_\beta = y_\beta$ for 
all $\beta \leq \alpha$, or there is some $\beta \leq \alpha$ with $x_\beta < y_\beta$.

Indeed, if $x_\alpha \in Y_\alpha$ for all $\alpha < \kappa$, then $x_\alpha = y_\alpha$ 
for all $\alpha < \kappa$. In the opposite case there is a least 
$\alpha$ with $x_\alpha \not \in Y_\alpha$. Then $\alpha > 0$, and $x_\beta \in Y_\beta$ 
for all $\beta < \alpha$. Hence, if $\beta < \alpha$,  then $x_\gamma = y_\gamma$ 
for all $\gamma < \beta$, showing that $\alpha$ is not a limit ordinal. Thus, $\alpha$ is 
successor ordinal, say $\alpha = \beta + 1$,  moreover, $x_\beta  \in Y_\beta$ and $x_\alpha \not\in Y_\alpha$. 
This implies that $x_\beta < y_\beta$ and $x_\gamma = y_\gamma$ for all $\gamma < \beta$, 
so that $x \sqsubset_\beta y$.

{\em Claim.} Let $z = (z_\alpha)_{\alpha < \kappa} \in L$ with $X \sqsubseteq z$ and $f(z) \sqsubseteq z$. 
Then for all $\alpha < \kappa$, either $y_\beta = z_\beta$ for all $\beta < \alpha$, 
or there is some $\beta \leq \alpha$ with $y_\beta < z_\beta$.

Indeed, suppose that $\alpha < \kappa$ and the claim holds for all ordinals less than $\alpha$. 
If $y_\beta < z_\beta$ for some $\beta < \alpha$ then we are done. 
Suppose now that $y_\beta = z_\beta$ for all $\beta < \alpha$. 
Then $f_\beta(z_\beta) = f_\beta(y_\beta) = y_\beta = z_\beta$
for all $\beta < \alpha$. Thus, if  $Y_\alpha$ is empty, 
then $y_\alpha$ is the least (pre-)fixed point of $f_\alpha$ in $Z_\alpha$,
whereas $z_\alpha$ is another pre-fixed point of $f_\alpha$ in $Z_\alpha$. 
Hence $y_\alpha \leq z_\alpha$. Suppose now that $Y_\alpha$ is not empty. 
Then $y_\alpha$ is the least pre-fixed point of $f_\alpha$ in $Z_\alpha$ 
above $Y_\alpha$, while $z_\alpha$ is another such pre-fixed point, 
since by $f(z) \sqsubseteq z$, $X \sqsubseteq z$ 
and $f_\beta(z_\beta) = z_\beta$ and $y_\beta = z_\beta$ for all $\beta < \alpha$ 
we have $f_\alpha(z_\alpha) \leq z_\alpha$ and $Y_\alpha \leq z_\alpha$. 
 We conclude that 
$y_\alpha \leq z_\alpha$.

It follows from the above claim that $y \sqsubseteq z$ whenever $X \sqsubseteq z$ 
and $f(z) \sqsubseteq z$.
\eop

By a similar argument, we can prove:

\begin{cor}
Let $L$ be a model satisfying the axioms A1--A6 and $f: L \to L$ be $\alpha$-monotone
for all $\alpha < \kappa$. Suppose that $X \subseteq L$ is a set of pre-fixed 
points of $f$ with resect to the ordering $\leq$. Then there is a (necessarily unique) 
$y \in L$ with the following properties:
\begin{itemize}
\item $y \sqsubseteq X$ and $f(y) = y$,
\item for all $z\in L$, if $z \sqsubseteq X$ and $z \sqsubseteq f(z)$, then 
      $z \sqsubseteq y$.
\end{itemize} 
\end{cor} 

{\sl Proof.} 
Again, we may assume that $L$ is a limit model. 
Using the notation introduced in the previous proof, for each $\alpha < \kappa$ define 
\begin{eqnarray*}
Y_\alpha&=& \{x \in X_\alpha : \forall \beta < \alpha \ h^\alpha_\beta(x) = y_\beta\}
\end{eqnarray*}
and let $y_\alpha$ be the greatest (post-)fixed point of $f_\alpha$ below $Y_\alpha$ in $Z_\alpha$, 
where $Z_\alpha$ is the set of all elements $z$ of $L_\alpha$ with $h^\alpha_\beta(z) = y_\beta$
for all $\beta < \alpha$.
Then $y = (y_\alpha)_{\alpha < \kappa}$ is the required element of $L$.
\eop  

\begin{cor}
\label{cor-latticefp}
Suppose that $L$ is a model and $f: L \to L$ is $\alpha$-monotone for all $\alpha < \kappa$.
Then the fixed points of $f$ form a complete lattice with respect to the ordering $\sqsubseteq$.
\end{cor}

\begin{cor}
\label{cor-lattice}
For every model $L$ satisfying the axioms A1--A6, $(L,\sqsubseteq)$ is a complete lattice. 
\end{cor}

{\sl Proof.} 
Let $f$ be the identity function in Corollary~\ref{cor-latticefp}. 
In particular, we obtain that if $X \subseteq L$, then the supremum 
$\bigsqcup X$ of $X$ w.r.t. the ordering $\sqsubseteq$ can be constructed as follows.
For each $\alpha < \kappa$, define 
\begin{eqnarray*}
Y_\alpha&=& \{x \in X_\alpha : \forall \beta < \alpha \ h^\alpha_\beta(x) = y_\beta\}
\end{eqnarray*}
and let $y_\alpha$ be the supremum of $Y_\alpha$ and the least element 
of $Z_\alpha$ in the complete lattice $L_\alpha$  (or in $Z_\alpha$). 
Then $\bigsqcup X = (y_\alpha)_{\alpha < \kappa}$. Note that if $Y_\alpha$ 
is empty, then $y_\alpha = \bigvee_{\alpha < \kappa} k^\alpha_\beta(y_\beta)$. 

The infimum $\bigsqcap X$ can be constructed dually. 
\eop

\begin{cor}
\label{cor-leastfp}
Let $L$ be a model satisfying the axioms A1--A6 and suppose that $f: L \to L$ 
is $\alpha$-monotone for all $\alpha < \kappa$. Then $f$ has a least pre-fixed 
point w.r.t. the ordering $\sqsubseteq$ which is a fixed point. 
Hence, if $x$ is the least fixed point of $f$ and $f(y) \sqsubseteq y$, then $x \sqsubseteq y$. 
\end{cor} 

{\sl Proof.} Let $X$ be the empty set in Theorem~\ref{thm-general}. \eop 

\begin{remark}
{\rm 
Suppose that $L$ is a model and $f: L \to L$ is $\alpha$-monotone for all $\alpha < \kappa$.
Let $x$ denote the least (pre-)fixed point of $f$ w.r.t. $\sqsubseteq$. If $f(z) \leq z$
for some $z \in L$, then also $f(z) \sqsubseteq z$, hence $x \sqsubseteq z$. 
}
\end{remark}

\begin{expl} 
\label{expl-logic program}
{\rm \cite{RW}
Suppose that $Z$ is a denumerable set of propositional variables and $P$
is an at most countably infinite propositional logic program over $Z$, 
possibly involving negation. Thus $P$ 
is a countable set of instructions of the form 
$z \leftarrow \ell_1 \wedge \cdots \wedge \ell_k$, where $z \in Z$ and $\ell _i$ 
is a literal for each $i$. Consider the model $L = V_\Omega^Z$,
defined in Section~\ref{sec1}, where $\Omega$ is the least uncountable ordinal. 
 Then $P$ induces a function $f_P: L \to L$
which maps an interpretation $I \in L$ to the 
interpretation $J = f_P(I)$ such that 
$J(z) = \bigvee_{z  \leftarrow \ell_1\wedge \cdots \wedge \ell_k \in P} (I(\ell_1) \wedge \cdots \wedge I(\ell_k))$,
where for a negative literal 
$\ell = \neg y$, $I(\ell) = T_{\alpha+1}$ if $I(y) = F_\alpha$,
$I(\ell) = F_{\alpha+1}$ if $I(y) = T_\alpha$, and $I(\ell) = 0$ if $I(y) = 0$. 
Then $f_P$ is $\alpha$-monotone for all $\alpha < \Omega$. The semantics of $P$ 
is defined as the least fixed point of $f_P$ w.r.t. $\sqsubseteq$. 
}
\end{expl} 

We end this section by giving an alternative proof of a result from \cite{ERwollic}. 

\begin{thm}
\label{thm-supppfp}
Suppose that $L$ is a model
satisfying A1--A6 and $f: L\to L$ is $\alpha$-monotone for each $\alpha < \kappa$. 
Let $X \subseteq L$ be a set of post-fixed points of $f$ w.r.t. the ordering $\leq$. 
Then $y = \bigsqcup X$ is also a post-fixed point of $f$ w.r.t. $\leq$. 
\end{thm}

{\sl Proof.} 
Suppose that $L$ is an inverse limit model as above, and let 
 $y  = (y_\alpha)_{\alpha < \kappa} = \bigsqcup X$. 
As before, let $f$ be determined by the family 
of conditionally monotone functions $f_\alpha$, ${\alpha < \kappa}$. 
We prove the following claim by induction on $\alpha < \kappa$:
Let $y_\beta \leq f_\beta(y_\beta)$ for all $\beta < \alpha$. Then $y_\alpha \leq f_\alpha(y_\alpha)$. 
We will use the notation in the proof 
of Corollary~\ref{cor-lattice}. 

Note that since $X$ is a set of post-fixed points of $f$ w.r.t. $\leq$, 
for each $\alpha < \kappa$, the $\alpha$-component of each element of $X$ 
is a post-fixed point of $f_\alpha$ with respect to the ordering of $L_\alpha$.

We consider two cases. Suppose first that $Y_\alpha \neq \emptyset$. Then $y_\alpha = \bigvee Y_\alpha$.
Since every element of $Y_\alpha$ is a post-fixed point of $f_\alpha$,  $y_\alpha$ is also 
a post-fixed point of $f_\alpha$.
Indeed, $y_\alpha = \bigvee Y_\alpha \leq \bigvee f_\alpha (Y_\alpha) \leq f_\alpha(\bigvee Y_\alpha) = 
f_\alpha(y_\alpha)$. Here, the second inequality is due to the fact that $f_\alpha$ 
is conditionally monotone and 
$h^\alpha_\beta(Y_\alpha) = \{y_\beta\}$ hence $h^\alpha_\beta(\bigvee Y_\alpha) = y_\beta$ for all 
$\beta < \alpha$.

Suppose next that $Y_\alpha = \emptyset$. Then $y_\alpha = \bigvee_{\beta < \alpha}  k^\alpha_\beta(y_\beta)$
is the least element of $Z_\alpha = \{z \in L_\alpha : 
\forall \beta < \alpha\ h^\alpha_\beta(z) = y_\beta\}$. 
Now for all $\beta < \alpha$, $y_\beta \leq f_\beta(y_\beta) = f(y)_\beta$, 
the $\beta$-component of $f(y)$. Thus,  $k^\alpha_\beta(y_\beta) \leq f(y)_\alpha$
for all $\beta < \alpha$, since $k^\alpha_\beta$ is an embedding. 
It follows that $y_\alpha \leq f(y)_\alpha = f_\alpha(y_\alpha)$. 
\eop 

We note that the dual also holds. If 
$L$ is a model and $f: L \to L$ is $\alpha$-monotone for all $\alpha < \kappa$, 
and if $X$ is set of pre-fixed points of $f$ w.r.t. the ordering  $\leq$,
then $\bigsqcap X$ is also a pre-fixed point.

\section{Symmetric models}
\label{sec-symmetric}

The first two axioms A1 and A2 and the axiom A6 introduced
in Section~\ref{sec1} are self dual, but the others are not. 

The dual forms of A3, A4 and A6 are given below. 
{\em 
\begin{itemize}
\item A3d. For all $x$ and $\alpha < \kappa$ there exists $y$ such that  $x=_\alpha y$
and for all $z$, if $z \sqsubseteq_\alpha  x$ then $z \leq y$. 
\end{itemize}
}
It is clear that $y$ is uniquely determined by $x$ and $\alpha$ and we will denote it by $x|^\alpha$.  
{\em 
\begin{itemize}
\item
A4d. For all $\alpha < \kappa$ and $x_i,y$, $i \in I$, where $I$ is 
a nonempty index set, if $x_i =_\alpha y$ for all $i \in I$, 
then $\bigwedge_{i \in I} x_i =_\alpha y$. 
\item
A5d. For all $x,y$ and $\alpha < \kappa$, if $x \leq y$ then $x|^\alpha \leq y|^\alpha$.
\end{itemize}
}
We also define the dual of A4$^*$.
{\em 
\begin{itemize}
\item
A4$^*$d. For all  $\alpha < \kappa$ and $x_i,y_i$ with 
    $x_i =_\alpha y_i$, $i \in I$, where $I$ is any index set,
    it holds that $\bigwedge_{i \in I}x_i =_\alpha \bigwedge_{i \in I} y_i$.
\end{itemize}
}

\begin{lem}
\label{lem-a3d}
There is a model not satisfying A3d. 
\end{lem}

{\sl Proof}. 
Consider the $4$-element lattice $(L,\leq)$ that is not a chain. Its elements are 
$\bot, 0,1,\top$ such that $\bot$ is least, $\top$ is greatest,
but there is no further nontrivial order relation. 

Define $\sqsubseteq_0$ to be the least preordering containing $\leq$ 
with respect to inclusion such that $\bot =_0 1$.
Let $\sqsubseteq_1$ be the least preordering with $\bot \sqsubseteq_1 1$, 
and for all $\alpha$ with $2 \leq \alpha < \kappa$, let $\sqsubseteq_\alpha$
be the identity relation. Then $L$ is a model but not a strong model:
A4$^*$ fails since $\bot =_0 1$ but $\bot \vee 0 = 0 \neq_0 \top = 1 \vee 0$. 
A3d fails since the set $\{ x : x \sqsubseteq_0 0\} = \{0,\bot,1\}$ has no greatest 
element w.r.t. $\leq$. Hence $0|^0$ does not exist. (Since A3d fails,
A5d makes no sense.) \eop



Regarding the dual of A4, the situation is different. 



\begin{lem}
\label{lem-a4dstar}
\label{lem-a4d}
Every model satisfying the axioms A1--A6 satisfies A4$^*$d. 
\end{lem}

{\sl Proof.}
Suppose that $L$ is a the model determined by the limit 
of an inverse system $h^\alpha_\beta : L_\alpha \to L_\beta$,
$\beta < \alpha < \kappa$ of complete lattices 
such that each $h^\alpha_\beta$ is a locally completely additive 
projection. 
Let $x_i,y_i\in L$ for all $i \in I$, and let $\alpha < \kappa$.
Suppose that $x_i \sqsubseteq_\alpha y_i$ for all $i \in I$. 
This means that for all $i \in I$, the $\alpha$-component of $x_i$ is 
less than or equal to the $\alpha$-component of $y_i$, and 
for all $\beta < \alpha$, the $\beta$-component of $x_i$ agrees with the 
$\beta$-component of $y_i$. Since the infimum is formed pointwise, it follows 
that the $\alpha$-component of $\bigwedge_{i \in I}x_i$ is less than or equal 
to the corresponding component of $\bigwedge_{i \in I} y_i$, 
whereas for all $\beta < \alpha$, the $\beta$-component of $\bigwedge_{i \in I}x_i$ is equal to the corresponding component of $\bigwedge_{i \in I} y_i$. \eop

\begin{lem}
\label{lem-stronga3d}
Every strong model satisfies A3d and A5d.
\end{lem} 

{\sl Proof.} Suppose that $L$ is a strong model. 
We use the Representation Theorem to prove that $L$ satisfies A3d.

So let $L$ be the model determined by the limit 
 of the inverse system of complete lattices $L_\alpha$, $\alpha < \kappa$,
with completely additive projections $h^\alpha_\beta : L_\alpha \to L_\beta$,
$\beta < \alpha < \kappa$. 
Let $x =(x_\gamma)_{\gamma < \kappa}$ in $L$ and $\alpha < \kappa$. 
Then let $x|^\alpha = \bigvee \{y  :  y \sqsubseteq_\alpha x\} = 
\bigvee \{ y : y_\alpha \leq x_\alpha\}$, where $y_\alpha$ 
denotes the $\alpha$-component of $y$.
Since the limit projection $h^\infty_\alpha$ is completely additive, 
$x|^\alpha$ is the $\leq$-greatest element $y$ of $L$ with
$y \sqsubseteq_\alpha x$.  
Moreover, $x|^\alpha =_\alpha x$, since $x \sqsubseteq_\alpha x$. 
This proves that A3d holds in $L$.

To prove that A5d holds as well, suppose that $x \leq x'$ in $L$, 
where $x = (x_\gamma)_{\gamma < \kappa}$ and $x' = (x'_\gamma)_{\gamma < \kappa}$.
Since $x \leq x'$, we have $x_\gamma \leq x'_\gamma$ for all $\gamma < \kappa$. 
Let $\alpha < \kappa$ and $y = (y_\gamma)_{\gamma < \kappa} \sqsubseteq_\alpha x$.
Then $y_\alpha \leq x_\alpha$ and $y_\beta = x_\beta$ for all $\beta < \alpha$.
Let $z = x' \vee y$. Since the functions $h^\infty_\gamma$ preserve suprema,
 we have $z_\gamma = x'_\gamma \vee y_\gamma$ for all $\gamma < \kappa$. 
In particular, $z_\gamma  = x'_\gamma$ for all $\gamma \leq  \alpha$,
proving $z \sqsubseteq_\alpha x'$. We have shown that for each $y\sqsubseteq_\alpha x$ there is 
some $z \sqsubseteq_\alpha x'$ with $y \leq z$. Thus, 
$x|^\alpha = \bigvee\{y : y \sqsubseteq_\alpha x\} \leq \bigvee\{z : z \sqsubseteq_\alpha x'\} = x'|^\alpha$.
\eop

Suppose that $L$ is a stratified complete lattice.
 We say that $L$ is a \emph{dual model} 
if it satisfies A1, A2, A3d, A4d, A5d and A6. Moreover
we call $L$ a \emph{strong dual model} if satisfies A1, A2, A3d, A4$^*$d, A5d and A6. 
Alternatively, $L$ is a (strong) dual model iff its dual $L^\op$, obtained 
by reversing the relation $\leq$ and each relation $\sqsubseteq_\alpha$,
is a (strong) model.

Of course, if a property holds in all models, then the dual property holds in
all dual models, and similarly for strong models. In particular, every (strong) dual 
model can be constructed as an inverse limit model. However, one uses 
dual projection-embedding pairs and locally infimum preserving or 
infimum preserving functions $h^\alpha_\beta: L_\alpha \to L_\beta$ 
of complete lattices. Here, when $L$ and $L'$ are complete lattices,
we say that $g: L'\to L$ is a dual projection with corresponding dual 
embedding $f: L \to L'$ if $f$ and $g$ are monotone, $g\circ f: L \to L$ is the identity
function on $L$, and $f \circ g : L'\to L'$ is greater than or 
equal to the identity function on $L'$. Alternatively, this means 
that $g$ is a projection $(L')^\op \to L^\op$ and $f$ is the corresponding
embedding $L^\op \to (L')^\op$. And a function $h: L' \to L$ 
is locally infimum preserving if for all $Y\subseteq L'$ and $x \in L$
with $h(Y) = x$, it holds that $h(\bigwedge Y) = x$. 
This clearly means that $h$ is locally completely additive
as a mapping of $(L')^\op$ into $L^\op$.

Every dual model is isomorphic to a model
determined by the limit of an inverse system 
$h^\alpha_\beta: L_\alpha \to L_\beta$ of 
locally infimum preserving dual projections. Moreover, 
every strong dual model is determined by such an inverse system 
where each $h^\alpha_\beta$ is a dual projection preserving all infima.
Dual models share several properties of models, e.g,.
each dual model $L$ gives rise to a complete lattice $(L,\sqsubseteq)$, 
and if $f: L \to L$ is $\alpha$-monotone for all $\alpha < \kappa$, where $L$ is a dual model,
then the set of all fixed points of $f$, ordered by 
$\sqsubseteq$, is a complete lattice.

We also define \emph{symmetric models} which are both models and dual models. 
Similarly, a \emph{strong symmetric model} is a strong model that is a 
strong dual model. 
As an immediate consequence of Lemma~\ref{lem-a4d} we have:
\begin{cor}
\label{cor-symmetric}
A model is symmetric iff it satisfies A3d and A5d.
\end{cor}

The standard model $V^Z$ discussed in Section~\ref{sec1} is a strong symmetric 
model as is any product model. But a  model may not 
be symmetric. See Lemma~\ref{lem-a3d}. 
Below we will show that the symmetric models are exactly the 
strong models, and in fact the strong symmetric models.

\begin{thm}
\label{thm-symmetric} 
The following conditions are equivalent for a model $L$ satisfying the axioms A1--A6. 
\begin{itemize}
\item $L$ is a strong model.
\item $L$ is a strong symmetric model.
\item $L$ is a symmetric model.
\end{itemize}
\end{thm}

{\sl Proof.} Suppose that $L$ is a strong model. Then $L$ is a symmetric model by 
Corollary~\ref{cor-symmetric} and Lemma~\ref{lem-stronga3d}. Suppose now that 
$L$ is a symmetric model. Then by Lemma~\ref{lem-a4dstar} and its dual, $L$ 
is a strong symmetric model. Finally, if $L$ is a strong symmetric model, then it is 
clearly a strong model. \eop

\begin{cor}
Let $L$ be a model determined by an inverse system of complete lattices 
$L_\alpha$, $\alpha < \kappa$, with locally completely additive projections 
$h^\alpha_\beta: L_\alpha \to L_\beta$. 
Then $L$ is a (strong) symmetric model iff the functions $h^\alpha_\beta$,
$\beta < \alpha < \kappa$  are completely 
additive.
\end{cor} 

Thus, in this case, the functions $h^\alpha_\beta$ preserve arbitrary infima and 
suprema.

\begin{cor}
A model is a (strong) symmetric model iff it is isomorphic to the model determined by 
an inverse system of complete lattices 
$L_\alpha$, $\alpha < \kappa$, with completely additive projections 
$h^\alpha_\beta: L_\alpha \to L_\beta$. 
\end{cor}

\section{Conclusion}

An axiomatic framework as an abstraction of the treatment of the semantics of logic programs 
with negation in \cite{RW} has recently been introduced in \cite{ERfp,ERwollic}. 
Here, we dealt with the models of two of the axiom systems of \cite{ERfp,ERwollic}, 
and established representation theorems for them. We proved that every 
model can be constructed from an inverse system of complete lattices with 
locally completely additive projections.
We also proved that every strong model can be constructed from an inverse system of complete 
lattices with completely additive projections. Using the inverse limit representation, 
we proved Theorem~\ref{thm-general} that asserts that the fixed points 
of a weakly monotone function over a model form a complete lattice with 
respect to a new ordering. In particular, there is a least fixed point, 
called the stratified least fixed point. 

We also studied models satisfying, together with each 
axiom, the dual axiom. We proved that such symmetric models are exactly 
the strong models, and in fact the strong symmetric models. In future work 
we intend to extend the representation theorem to more general classes of models 
introduced in \cite{ERfp}, where the preorderings $\sqsubseteq_\alpha$ are not 
completely determined by the ordering $\leq$ and the equivalence relations $=_\alpha$.    

Since the semantics of recursive definitions is usually 
captured by fixed points of functions, or functors, or other 
constructors, fixed point operations 
appear in almost all branches of computer
science including automata and languages, semantics, concurrency, programming 
logics, the characterization of complexity classes using 
formal logic, etc. Among the prominent fixed point theorems commonly used 
in computer science are
the least fixed point theorem of Knaster and Tarski 
and the fixed point theorem of Kleene, that apply to monotone or order continuous 
functions over complete lattices or cpo's, see \cite{Daveyetal,Tarski},
or their categorical generalizations \cite{AdamekKoubek,Lambek,PlotkinSmyth,Wand}, 
or in a metric setting, the Banach fixed point theorem \cite{Banach}. 
It has been shown for each that the corresponding fixed point operation
satisfies the same equational laws, captured by the notion of iteration 
theories \cite{BEbook,EsMFCS}. 

Our aim with this paper and its predecessors has been to contribute to the 
development of a novel general framework for solving fixed point equations involving 
non-monotone operations as an alternative of the bilattice based approach 
\cite{Deneckeretal1,Deneckeretal,Fitting,Prz}. This method has already found applications 
in logic programming and Boolean context-free grammars,
 and we plan to apply it in other situations
including Boolean automata, fuzzy sets, and quantitative logics. 
A nice feature of the approach is that the 
stratified least fixed point operation over weakly monotonic
functions also satisfies the standard equational laws, 
cf. \cite{Esax}.

\end{document}